\documentclass[12pt]{article}
\usepackage{amsmath,axodraw,cite}
\numberwithin{equation}{section}
\newcommand{\Tr}{\mathop{\rm Tr}\nolimits}

\def\slash#1{\setbox0=\hbox{$#1$}               
   \dimen0=\wd0                                 
   \setbox1=\hbox{/} \dimen1=\wd1               
   \ifdim\dimen0>\dimen1                        
      \rlap{\hbox to \dimen0{\hfil/\hfil}}      
      #1                                        
   \else                                        
      \rlap{\hbox to \dimen1{\hfil$#1$\hfil}}   
      /                                         
   \fi}                                         %

\newcommand{\eqnoffst}{-1.3 cm}
\newcommand{\eqnoffsta}{-1.6 cm}
\newcommand{\eqnoffstc}{-2.3cm}
\newcommand{\eqnoffstb}{-0.0cm}

\begin{document}

\begin{titlepage}

\begin{flushright}
MPI-PhT/98-91\\
March 1999\\
\end{flushright}
\vskip 2cm
\begin{center}
{\Large \bf Complete ${\cal O}(N_f\alpha^2)$ Weak Contributions\\
            to the Muon Lifetime} \\[20mm]
  Paresh Malde and Robin G. Stuart \\ [2mm]
\vskip 0.5cm {\it Randall Physics Laboratory, University of Michigan\\
                  Ann Arbor, MI 48109-1120, USA}\\
\end{center}
\vskip 3cm
\hrule
\begin{abstract}
The complete ${\cal O}(N_f\alpha^2)$ weak contributions to the muon
lifetime, denoted as $\Delta r^{(2)}$,
are calculated in the $\overline{{\rm MS}}$ renormalization
scheme. These come from 2-loop Feynman diagrams containing a loop
formed by complete generations of massless fermions.
They form an independent, gauge-invariant
set of corrections that, because of the large number of light fermions in the
Standard Model, is expected to make a significant contribution.
In the $\overline{{\rm MS}}$ renormalization scheme with
$\mu^\prime\equiv(\pi e^\gamma)^{\frac{1}{2}}\mu=M_Z$
and for a Higgs mass, $M_H$, in the range 100--1000\,GeV the contribution
varies from $-0.55\times 10^{-4}$ to $-1.54\times 10^{-4}$
for each massless generation of fermions.
\end{abstract}
\hrule
\vspace{2mm}

\end{titlepage}

\setcounter{footnote}{0} \setcounter{page}{2} \setcounter{section}{0}
\newpage

\section{Introduction}

The Fermi coupling constant, $G_F$, is extracted from the measured value
of the muon lifetime, $\tau_\mu$ via the formula
\begin{equation}
\frac{1}{\tau_\mu}\equiv\Gamma_\mu=\Gamma_0(1+\Delta q).
\label{eq:QEDcorr}
\end{equation}
where $\Delta q$ encapsulates the radiative corrections to all orders in
$\alpha$ as calculated using the lagrangian of the Fermi theory
\cite{KinoSirl,muonhad,muonprl,muon1ppm}.
This value of $G_F$ is then related to the
parameters of the Standard Model of electroweak interactions by
\begin{equation}
\frac{G_F}{\sqrt{2}}=\frac{g^2}{8M_W^2}(1+\Delta r)
\end{equation}
in which $g$ and $M_W$ are the renormalized $SU(2)_L$ coupling constant
and $W$ boson mass, respectively, in whatever renormalization scheme
has been chosen.

The quantity $\Delta r$ was introduced by Sirlin \cite{Sirlin80} and is
intimately related to the $\rho$-parameter \cite{RossVeltman}
\begin{equation}
\rho=\frac{M_W^2}{M_Z^2\cos^2\theta_W}=1+\delta\rho
\end{equation}
where $\theta_W$ is the weak mixing angle.
For many classes of radiative corrections, generally those that appear
only in the $W$ and $Z^0$ self-energy diagrams, $\delta\rho$ is
related to $\Delta r$ by the simple relation,
\begin{equation}
\delta\rho=-\frac{\sin^2\theta_W}{\cos^2\theta_W}\Delta r
\end{equation}
Since $\delta\rho$ was constructed as a way of interrogating
the Higgs or mass generation sector of
of the theory it plays a central r\^ole in our quest to understand this
largely unexplored feature of the Standard Model.
This endeavour has already borne fruit. It was the inordinately strong
dependence of $\delta\rho$ on the top quark mass,
\begin{equation}
\delta\rho\sim\left(\frac{\alpha}{4\pi}\right)
                    \frac{3m_t^2}{4M_W^2\sin^2\theta_W},
\end{equation}
that allowed its
value to be predicted from precision electroweak data before it
was directly observed at the TeVatron \cite{CDF}.
For this reason a great deal of effort has been
devoted to calculating classes of 2-loop electroweak corrections
contributing to $\Delta r$
\cite{BijVelt,Frank,ConsHollJege,Barbieri1,Barbieri2,Fleischer1,
Fleischer2,DegrFancGamb,BaubWeig}.

Here the ${\cal O}(N_f\alpha^2)$ corrections to $\Delta r$ are given.
These are the 2-loop corrections containing a massless fermion loop.
Since the number of fermions, $N_f$, is quite large this class of
corrections can be reasonably expected to constitute a dominant subclass.
Moreover the scaling with $N_f$ provides a unique tag and the complete set of
corrections will therefore be gauge-invariant.

This type of enhancement is already seen in the decay widths of the $W$
and $Z^0$ bosons that are much broader than typical weak resonances
due dominantly to
the large number of decay channels available to them. The
majority of the ${\cal O}(N_f\alpha^2)$ diagrams that occur in
$\Delta r$ contain the same multiplicative factor, i.e.\ squares
of coupling constants summed over light
fermion species, that are responsible
for broadening the weak vector bosons.

${\cal O}(N_f\alpha^2)$ corrections have been discussed elsewhere
\cite{WeigSchaBohm} for the self-energy diagrams of the $W$ and
$Z^0$ at general $q^2$. In the present calculation,
Feynman diagrams need only
be evaluated at $q^2=0$ which simplifies matters considerably
and yields much more compact and tractable results.
However other classes of diagrams, vertex and box diagrams, now arise.
Moreover the complete ${\cal O}(N_f\alpha^2)$ renormalization must be
confronted. A detailed study of renormalization at this order was
carried out in the context of electric charge renormalization in
ref.\cite{MaldeStuart1}. In that case the unbroken $U(1)$ symmetry
generates a large number of interrelationships that can be used to test
consistency of the renormalization procedure.

In section \ref{eq:NoteConv} the notation and conventions used
are set out in detail. In general these are identical to those adopted
in ref.\cite{MaldeStuart1}.
In section 3 a brief discussion of Ward identities is given, reviewing
the lessons learned from ref.\cite{MaldeStuart1} and how they are to be
applied to in the present calculation. In section~4 the
${\cal O}(N_f\alpha^2)$ corrections to $\Delta r^{(2)}$ are given
separately for the self-energy, vertex and box contributions.
The conclusions arising from the complete analytic result are given in
section~5. The appendices contain identities that were used in the
course of the calculation along with a complete list of the results
for all Feynman diagrams in terms of a single master integral.

\section{Notation and Conventions}
\label{eq:NoteConv}

\subsection{Renormalization}

In order to make a physical prediction the complete renormalization
of the Standard Model at ${\cal O}(N_f\alpha^2)$ must be carried out.
Renormalization at this order has been discussed in detail for a
general renormalization scheme in ref.\cite{MaldeStuart1} and
electromagnetic charge renormalization is considered in particular.
This allowed the divergent parts of the $W$-fermion vertex to be
predicted as explained in section 3.2. The notation adopted here
comes directly from ref.\cite{MaldeStuart1}.

Standard rescalings of the bare $SU(2)_L$ and $U(1)$ fields,
$W^0$ and $B^0$, are carried out to obtain their associated renormalized
fields, $W$ and $B$ and wavefunction counterterms, $\delta Z_W$ and
$\delta Z_B$. The bare $SU(2)_L$ and $U(1)$ coupling constants,
$g^0$ and $g^{\prime0}$, are treated similarly as are the squares of the
bare $W$ and $Z$ boson masses, $(M_W^2)^0$ and $(M_Z^2)^0$. Thus
\begin{alignat}{3}
W^0&=(1+\delta Z_W)^{\frac{1}{2}}W&\qquad\qquad
g^0&=g+\delta g&\qquad\qquad
(M_W^2)^0&=M_W^2+\delta M_W^2\label{eq:renorm1}\\
B^0&=(1+\delta Z_B)^{\frac{1}{2}}B&\qquad\qquad
g^{\prime0}&=g^\prime+\delta g^\prime&\qquad\qquad
(M_Z^2)^0&=M_Z^2+\delta M_Z^2\label{eq:renorm2}
\end{alignat}

Expressions for the required counterterms in the charged sector of the
theory generated by the substitutions (\ref{eq:renorm1}) and
(\ref{eq:renorm2}) are given later in this paper and those for the
neutral sector can be found in ref.\cite{MaldeStuart1}.

The weak mixing angle, $\theta_W$, is defined so as to diagonalize the
mass matrix of the neutral $W_3$ and $B$ fields in the renormalized
lagrangian. $s_\theta$ and $c_\theta$ are used to denote the sine and
cosine of $\theta_W$ respectively. The relation
$c_\theta^2=M_W^2/M_Z^2$ then
holds exactly in any renormalization scheme provided $M_W$ and
$M_Z$ are the renormalized masses in the particular renormalization
scheme being used.

It is important to note that this choice for $\theta_W$ is not the only
possibility. The weak mixing angle can also be defined so as
diagonalize the
mass matrix of the bare lagrangian but this will then generate a
counterterm, $\delta\theta_W$, which is inconvenient and unnecessarily
complicated in practice.

In the present calculation it is necessary
to distinguish between 1-loop fermionic and bosonic
corrections. The order and type of a correction will be indicated,
by a superscript in parentheses. Thus $\delta Z^{(1f)}$
indicates the 1-loop fermionic part of the counterterm $\delta Z$.
The 1-loop bosonic corrections are denoted by the superscript ${}^{(1b)}$
and the superscript ${}^{(1)}$ indicates both together. The superscript
${}^{(2)}$ when used here means the full ${\cal O}(N_f\alpha^2)$ correction.

Throughout this work the Euclidean metric is used with the square of
time-like momenta being negative. The calculation is performed in
't~Hooft-Feynman, $R_{\xi=1}$, gauge.

A fully anti-commuting Dirac $\gamma_5$ will be assumed.
This could only lead to difficulties in fermion loops that generate
the antisymmetric $\epsilon$ tensor, such as internal fermion triangles.
Anomaly cancellation in the sum over a complete generation
guarantees that additional terms cannot appear. Care must also be taken
in the case of external fermion currents where three $\gamma$ matrices
come together. This occurs for the case of box diagrams and will be
discussed further in section~4.3.

It was shown in ref.\cite{loweng} that all ${\cal O}(N_f\alpha^2)$
Feynman diagrams contributing to $\Delta r^{(2)}$ can be reduced to
expressions in terms of the single master integral given in
Appendix~\ref{sect:MasterInt}.

The calculation was performed in a general renormalization scheme but
only results for $\overline{{\rm MS}}$ are presented here.
Expressions in this scheme are generally much more compact
since the finite parts of counterterms are absent.
In addition, for massless fermions, the ${\cal O}(N_f\alpha^2)$
vector-scalar mixing and 2-point scalar counterterms, that could be
finite in a general renormalization scheme, vanish.
Most of the counterterms required for the calculation of muon decay
can be obtained by evaluating 2-loop Feynman diagrams at momentum
$q^2=0$. An exception to this is the $W$ boson mass counterterm that is
obtained from the $W$ boson self energy evaluated at $q^2=-M_W^2$. In
the $\overline{{\rm MS}}$ renormalization scheme only the divergent
parts are required, however, and these are much easier to calculate than
the finite parts.

Renormalization schemes differ only in the finite parts of their
counterterms and the so-called on-shell scheme \cite{Sirlin80}
is the most widely-used alternative to $\overline{{\rm MS}}$
in electroweak physics.
Explicit expressions for all ${\cal O}(N_f\alpha^2)$ Feynman diagrams
that occur in the calculation are given in the appendices. These,
of course, do not depend on the renormalization scheme provided
the parameters, $g$, $s_\theta$, $c_\theta$, $M_W$ and $M_Z$ are
interpreted as renormalized parameters in the particular scheme being
used. General expressions for the complete set of 1-loop diagrams
containing 1-loop counterterms, which are formally of
${\cal O}(N_f\alpha^2)$, are straightforward if tedious to obtain.
They are rather lengthy and so are not given here.
Appendix~\ref{sect:CTInsert} contains
some identities that are useful for the calculation of this class of
diagrams where there are counterterm insertions on internal
photons and $Z^0$'s.

The ${\cal O}(N_f\alpha^2)$ corrections to the muon lifetime
in the $\overline{{\rm MS}}$ renormalization scheme
constitute one of those rare cases where 2-loop electroweak corrections can be usefully
written down. In most other cases the expressions are sufficiently
complex that there is little point for them to exist outside of a
computer program.

\subsection{The $\overline{{\rm MS}}$ renormalization scheme}

The $\overline{{\rm MS}}$ renormalization scheme is implemented in
dimensional regularization by requiring that the counterterms contain
only pole pieces obtained by Laurent expansion of divergent quantities
about $n=4$ as they would be under minimal subtraction, MS. Thus a
general 2-loop counterterm takes the form,
$\delta Z^{(2)}=a_{-2}\epsilon^{-2}+a_{-1}\epsilon^{-1}$
where $\epsilon=2-n/2$ and $a_{-2}$ and $a_{-1}$ are constants.
In addition the 't~Hooft mass, $\mu$, is written in terms of the
rescaled $\mu^\prime$ with
$\mu=(\pi e^\gamma)^{-\frac{1}{2}}\mu^\prime$.
This has the effect of eliminating many of the uninteresting constants
that occur at intermediate stages. At 1-loop order this procedure is
equivalent to defining the counterterms as being proportional to
$\Delta=\epsilon^{-1}-\gamma-\ln\pi$ but without rescaling
the 't~Hooft mass.

The exact value that is chosen for $\mu^\prime$ depends on the
particular application that is being considered. For the analysis of
electroweak data obtained around the $Z^0$ resonance
$\mu^\prime=M_Z$ is a reasonable choice since it eliminates the need to
resum large logarithms associated with the running of the
electromagnetic coupling constant.

\section{Ward identities}

In ref.\cite{MaldeStuart1} the ${\cal O}(N_f\alpha^2)$ renormalization
of the Standard Model was studied in detail following the prescription
of Ross and Taylor \cite{RossTaylor}. There the gauge-fixing lagrangian
is constructed from renormalized, rather than bare, fields in order to
satisfy the Ward identities of the theory with the result that the mixing
between the vector bosons and Goldstone scalars is no longer completely
canceled in $R_\xi$ gauges. It can be shown that this is formally
equivalent to schemes where the gauge parameter is renormalized
\cite{BohmHollSpie,Hollik}
but is often more convenient to apply in practice.
The ${\cal O}(N_f\alpha^2)$ counterterms that mix vectors and scalars
are finite in a general renormalization scheme and therefore vanish in
the $\overline{{\rm MS}}$ scheme.

Renormalization of the Standard Model has been exhaustively studied at
the 1-loop level. There is some flexibility as to whether wavefunction
counterterms, $\delta Z^{(1)}$ are used or not. If they are employed the
Green's functions are rendered finite but the wavefunction counterterms
cancel out when the Green's functions
are combined to form physical $S$-matrix elements. This is demonstrated
in Appendix~\ref{sect:CTInsert}. It follows that the
wavefunction counterterms can be dropped altogether,
for example see ref.\cite{Sirlin80}, provided one is prepared to deal
with divergent Green's functions. The divergences will then cancel out
in overall physical matrix elements. In practice this feature provides
a useful check of the calculation.

In ref.\cite{MaldeStuart1} it was shown that at 2-loop order the
wavefunction counterterms, $\delta Z^{(2)}$ cancel in physical
matrix elements, and so can be dropped if desired, but that the
1-loop wavefunction counterterms must be included in a manner
consistent with the 1-loop Ward identities
\begin{eqnarray}
\frac{1}{2}\delta Z_B^{(1)}+\frac{\delta g^{\prime(1)}}{g^\prime}&=&0
\label{eq:WardidB}\\
\frac{1}{2}\delta Z_W^{(1f)}+\frac{\delta g^{(1f)}}{g}&=&0
\label{eq:WardidW}
\end{eqnarray}
where eq.(\ref{eq:WardidB}) is true for both fermionic and bosonic
counterterms separately.

The imposition of the Ward identities, (\ref{eq:WardidB}) and
(\ref{eq:WardidW}), leads to some quite substantial simplifications.
Further simplification can also be obtained by noting
that in any renormalization scheme
\begin{eqnarray}
\frac{1}{2}\delta Z_W^{(1b)}+\frac{\delta g^{(1b)}}{g}
+2\left(\frac{g^2}{16\pi^2}\right)(\pi M_W^2)^{-\epsilon}\Gamma(\epsilon)
&=&\mbox{finite}
\label{eq:finitecombo1}\\
\delta Z^{(1f)}_\phi=
\frac{\delta M_W^{2(1f)}}{M_W^2}-2\frac{\delta g^{(1f)}}{g}&=&\mbox{finite}
\label{eq:finitecombo2}
\end{eqnarray}
The former vanishes in the on-shell renormalization scheme and the latter
in $\overline{\rm MS}$. Here $\delta Z_\phi$ is the wavefunction
counterterm for the Higgs field.

\section{The ${\cal O}(N_f\alpha^2)$ Corrections to $\Delta r^{(2)}$}

\subsection{The $W$-boson Self-Energy}
\label{sect:WSelfe}

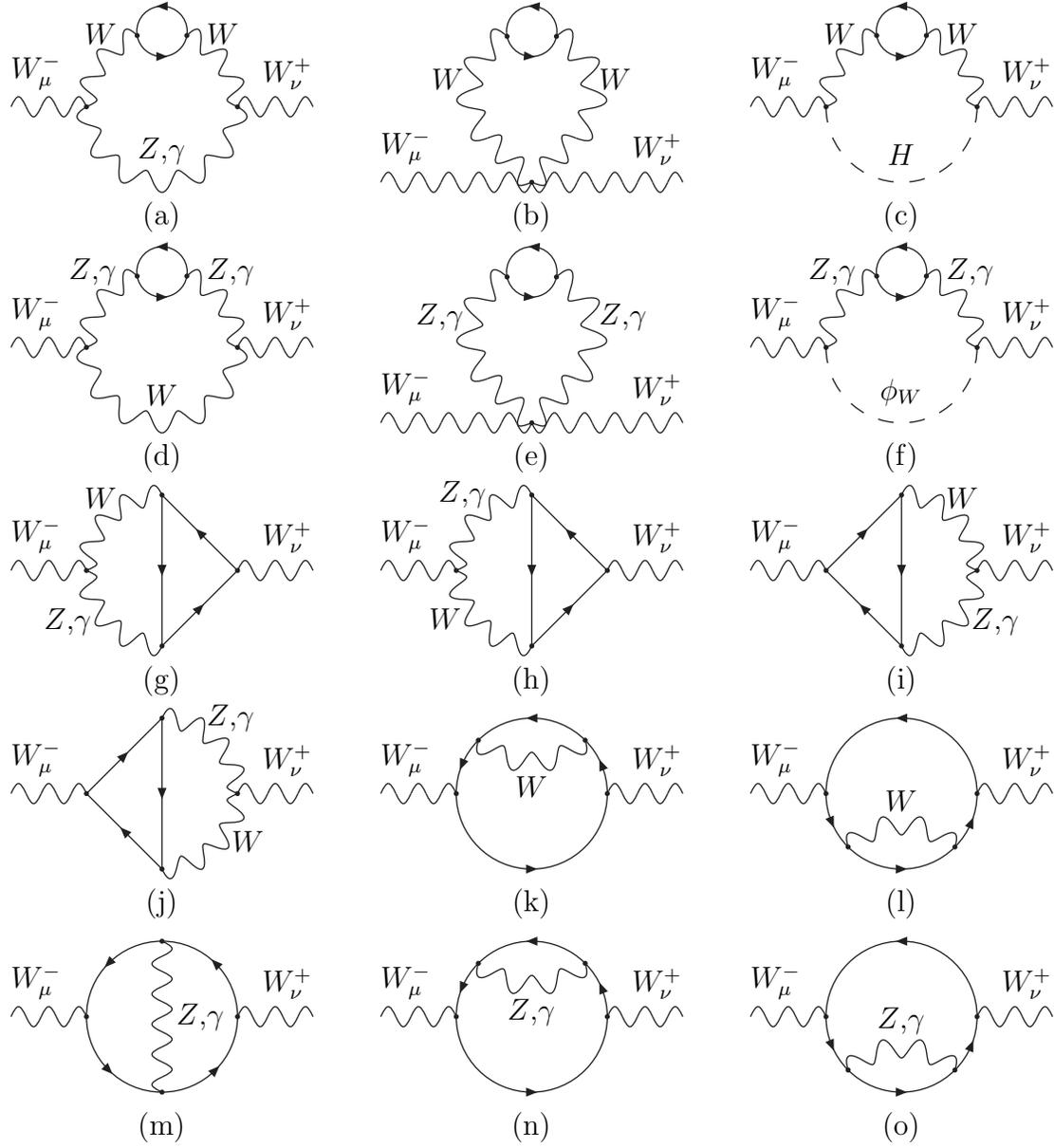
\begin{figure}
\begin{center}
\begin{picture}(120,80)(0,-45)
\Photon(0,0)(30,0){4}{2.5}        \Vertex(30,0){1}
\Text(0,7)[bl]{$W^-_\mu$}
\PhotonArc(60,0)(30,0,70.81){-4}{3}
\Text(42,25)[br]{$W$}
\PhotonArc(60,0)(30,109.18,180){4}{3}
\Text(78,25)[bl]{$W$}
\PhotonArc(60,0)(30,180,0){4}{6.5}
\Text(60,-23)[b]{$Z$,$\gamma$}
\Photon(90,0)(120,0){4}{2.5}      \Vertex(90,0){1}
\Text(120,7)[br]{$W^+_\nu$}
\ArrowArc(60,30)(10,-9.6,189.6)   \Vertex(50.14,28.33){1}
\ArrowArc(60,30)(10,189.6,-9.6)   \Vertex(69.86,28.33){1}
\Text(60,-38)[t]{(a)}
\end{picture}
\qquad
\begin{picture}(120,80)(0,-45)
\Photon(0,-30)(60,-30){4}{5}    \Vertex(60,-30){1}
\Photon(60,-30)(120,-30){-4}{5}
\Text(0,-23)[bl]{$W^-_\mu$}   \Text(120,-23)[br]{$W^+_\nu$}
\PhotonArc(60,5.15)(24.85,-45,66.78){5}{3.5}     \Vertex(50.20,27.99){1}
\PhotonArc(60,5.15)(24.85,113.22,225){5}{3.5}    \Vertex(69.80,27.99){1}
\Photon(60,-30)(42.43,-12.43){4.5}{2}
\Photon(60,-30)(77.57,-12.43){-4.5}{2}
\Text(33,7)[br]{$W$}
\Text(87,7)[bl]{$W$}
\ArrowArc(60,30)(10,-11.61,181.61)
\ArrowArc(60,30)(10,181.61,-11.61)
\Text(60,-38)[t]{(b)}
\end{picture}
\qquad
\begin{picture}(120,80)(0,-45)
\Photon(0,0)(30,0){4}{2.5}        \Vertex(30,0){1}
\Text(0,7)[bl]{$W^-_\mu$}
\PhotonArc(60,0)(30,0,70.81){-4}{3}
\Text(42,25)[br]{$W$}
\PhotonArc(60,0)(30,109.18,180){4}{3}
\Text(78,25)[bl]{$W$}
\DashCArc(60,0)(30,180,0){6}
\Text(60,-23)[b]{$H$}
\Photon(90,0)(120,0){4}{2.5}      \Vertex(90,0){1}
\Text(120,7)[br]{$W^+_\nu$}
\ArrowArc(60,30)(10,-9.6,189.6)   \Vertex(50.14,28.33){1}
\ArrowArc(60,30)(10,189.6,-9.6)   \Vertex(69.86,28.33){1}
\Text(60,-38)[t]{(c)}
\end{picture}
\\\vglue 15 pt
\begin{picture}(120,80)(0,-45)
\Photon(0,0)(30,0){4}{2.5}        \Vertex(30,0){1}
\Text(0,7)[bl]{$W^-_\mu$}
\PhotonArc(60,0)(30,0,70.81){-4}{3}
\Text(42,25)[br]{$Z$,$\gamma$}
\PhotonArc(60,0)(30,109.18,180){4}{3}
\Text(78,25)[bl]{$Z$,$\gamma$}
\PhotonArc(60,0)(30,180,0){4}{6.5}
\Text(60,-23)[b]{$W$}
\Photon(90,0)(120,0){4}{2.5}      \Vertex(90,0){1}
\Text(120,7)[br]{$W^+_\nu$}
\ArrowArc(60,30)(10,-9.6,189.6)   \Vertex(50.14,28.33){1}
\ArrowArc(60,30)(10,189.6,-9.6)   \Vertex(69.86,28.33){1}
\Text(60,-38)[t]{(d)}
\end{picture}
\qquad
\begin{picture}(120,80)(0,-45)
\Photon(0,-30)(60,-30){4}{5}    \Vertex(60,-30){1}
\Photon(60,-30)(120,-30){-4}{5}
\Text(0,-23)[bl]{$W^-_\mu$}   \Text(120,-23)[br]{$W^+_\nu$}
\PhotonArc(60,5.15)(24.85,-45,66.78){5}{3.5}     \Vertex(50.20,27.99){1}
\PhotonArc(60,5.15)(24.85,113.22,225){5}{3.5}    \Vertex(69.80,27.99){1}
\Photon(60,-30)(42.43,-12.43){4.5}{2}
\Photon(60,-30)(77.57,-12.43){-4.5}{2}
\Text(33,7)[br]{$Z$,$\gamma$}
\Text(87,7)[bl]{$Z$,$\gamma$}
\ArrowArc(60,30)(10,-11.61,181.61)
\ArrowArc(60,30)(10,181.61,-11.61)
\Text(60,-38)[t]{(e)}
\end{picture}
\qquad
\begin{picture}(120,80)(0,-45)
\Photon(0,0)(30,0){4}{2.5}        \Vertex(30,0){1}
\Text(0,7)[bl]{$W^-_\mu$}
\PhotonArc(60,0)(30,0,70.81){-4}{3}
\Text(42,25)[br]{$Z$,$\gamma$}
\PhotonArc(60,0)(30,109.18,180){4}{3}
\Text(78,25)[bl]{$Z$,$\gamma$}
\DashCArc(60,0)(30,180,0){6}
\Text(60,-23)[b]{$\phi_W$}
\Photon(90,0)(120,0){4}{2.5}      \Vertex(90,0){1}
\Text(120,7)[br]{$W^+_\nu$}
\ArrowArc(60,30)(10,-9.6,189.6)   \Vertex(50.14,28.33){1}
\ArrowArc(60,30)(10,189.6,-9.6)   \Vertex(69.86,28.33){1}
\Text(60,-38)[t]{(f)}
\end{picture}
\\\vglue 8 pt
\begin{picture}(120,80)(0,-45)
\Photon(0,0)(30,0){4}{2.5}         \Vertex(30,0){1}
\Text(0,7)[bl]{$W^-_\mu$}
\PhotonArc(60,0)(30,90,180){4}{4}
\Text(42,25)[br]{$W$}
\PhotonArc(60,0)(30,180,270){-4}{4}
\Text(32,-15)[tr]{$Z$,$\gamma$}
\ArrowLine(90,0)(60,30)           \Vertex(60,30){1}
\ArrowLine(60,30)(60,-30)         \Vertex(60,-30){1}
\ArrowLine(60,-30)(90,0)          \Vertex(90,0){1}
\Photon(90,0)(120,0){4}{2.5}
\Text(120,7)[br]{$W^+_\nu$}
\Text(60,-38)[t]{(g)}
\end{picture}
\qquad
\begin{picture}(120,80)(0,-45)
\Photon(0,0)(30,0){4}{2.5}         \Vertex(30,0){1}
\Text(0,7)[bl]{$W^-_\mu$}
\PhotonArc(60,0)(30,90,180){4}{4}
\Text(42,25)[br]{$Z$,$\gamma$}
\PhotonArc(60,0)(30,180,270){-4}{4}
\Text(32,-15)[tr]{$W$}
\ArrowLine(90,0)(60,30)           \Vertex(60,30){1}
\ArrowLine(60,30)(60,-30)         \Vertex(60,-30){1}
\ArrowLine(60,-30)(90,0)          \Vertex(90,0){1}
\Photon(90,0)(120,0){4}{2.5}
\Text(120,7)[br]{$W^+_\nu$}
\Text(60,-38)[t]{(h)}
\end{picture}
\qquad
\begin{picture}(120,80)(0,-45)
\Photon(0,0)(30,0){4}{2.5}
\Text(0,7)[bl]{$W^-_\mu$}
\PhotonArc(60,0)(30,0,90){-4}{4}
\Text(78,25)[bl]{$W$}
\PhotonArc(60,0)(30,270,0){4}{4}
\Text(88,-15)[tl]{$Z$,$\gamma$}
\ArrowLine(30,0)(60,30)           \Vertex(60,30){1}
\ArrowLine(60,30)(60,-30)         \Vertex(60,-30){1}
\ArrowLine(60,-30)(30,0)          \Vertex(30,0){1}
\Photon(90,0)(120,0){4}{2.5}      \Vertex(90,0){1}
\Text(120,7)[br]{$W^+_\nu$}
\Text(60,-38)[t]{(i)}
\end{picture}
\\\vglue 8 pt
\begin{picture}(120,80)(0,-45)
\Photon(0,0)(30,0){4}{2.5}
\Text(0,7)[bl]{$W^-_\mu$}
\PhotonArc(60,0)(30,0,90){-4}{4}
\Text(78,25)[bl]{$Z$,$\gamma$}
\PhotonArc(60,0)(30,270,0){4}{4}
\Text(88,-15)[tl]{$W$}
\ArrowLine(30,0)(60,30)           \Vertex(60,30){1}
\ArrowLine(60,30)(60,-30)         \Vertex(60,-30){1}
\ArrowLine(60,-30)(30,0)          \Vertex(30,0){1}
\Photon(90,0)(120,0){4}{2.5}      \Vertex(90,0){1}
\Text(120,7)[br]{$W^+_\nu$}
\Text(60,-38)[t]{(j)}
\end{picture}
\qquad
\begin{picture}(120,80)(0,-45)
\Photon(0,0)(30,0){4}{2.5}
\Text(0,7)[bl]{$W^-_\mu$}
\ArrowArc(60,0)(30,0,45)         \Vertex(81.21,21.21){1}
\ArrowArc(60,0)(30,45,135)       \Vertex(38.79,21.21){1}
\ArrowArc(60,0)(30,135,180)      \Vertex(30,0){1}
\ArrowArc(60,0)(30,180,0)        \Vertex(90,0){1}
\PhotonArc(60,42.42)(30,225,315){4}{3.5}
\Text(60,7)[t]{$W$}
\Photon(90,0)(120,0){4}{2.5}
\Text(120,7)[br]{$W^+_\nu$}
\Text(60,-38)[t]{(k)}
\end{picture}
\qquad
\begin{picture}(120,80)(0,-45)
\Photon(0,0)(30,0){4}{2.5}
\Text(0,7)[bl]{$W^-_\mu$}
\ArrowArc(60,0)(30,0,180)        \Vertex(30,0){1}
\ArrowArc(60,0)(30,180,225)      \Vertex(38.79,-21.21){1}
\ArrowArc(60,0)(30,225,315)      \Vertex(81.21,-21.21){1}
\ArrowArc(60,0)(30,315,0)        \Vertex(90,0){1}
\PhotonArc(60,-42.42)(30,45,135){4}{3.5}
\Text(60,-7)[b]{$W$}
\Photon(90,0)(120,0){4}{2.5}
\Text(120,7)[br]{$W^+_\nu$}
\Text(60,-38)[t]{(l)}
\end{picture}
\\\vglue 8 pt
\begin{picture}(120,80)(0,-45)
\Photon(0,0)(30,0){4}{2.5}
\Text(0,7)[bl]{$W^-_\mu$}
\ArrowArc(60,0)(30,0,90)         \Vertex(60,30){1}
\ArrowArc(60,0)(30,90,180)       \Vertex(30,0){1}
\ArrowArc(60,0)(30,180,270)      \Vertex(60,-30){1}
\ArrowArc(60,0)(30,270,0)        \Vertex(90,0){1}
\Photon(60,30)(60,-30){4}{5}
\Text(66,0)[l]{$Z$,$\gamma$}
\Photon(90,0)(120,0){4}{2.5}
\Text(120,7)[br]{$W^+_\nu$}
\Text(60,-38)[t]{(m)}
\end{picture}
\qquad
\begin{picture}(120,80)(0,-45)
\Photon(0,0)(30,0){4}{2.5}
\Text(0,7)[bl]{$W^-_\mu$}
\ArrowArc(60,0)(30,0,45)         \Vertex(81.21,21.21){1}
\ArrowArc(60,0)(30,45,135)       \Vertex(38.79,21.21){1}
\ArrowArc(60,0)(30,135,180)      \Vertex(30,0){1}
\ArrowArc(60,0)(30,180,0)        \Vertex(90,0){1}
\PhotonArc(60,42.42)(30,225,315){4}{3.5}
\Text(60,7)[t]{$Z$,$\gamma$}
\Photon(90,0)(120,0){4}{2.5}
\Text(120,7)[br]{$W^+_\nu$}
\Text(60,-38)[t]{(n)}
\end{picture}
\qquad
\begin{picture}(120,80)(0,-45)
\Photon(0,0)(30,0){4}{2.5}
\Text(0,7)[bl]{$W^-_\mu$}
\ArrowArc(60,0)(30,0,180)        \Vertex(30,0){1}
\ArrowArc(60,0)(30,180,225)      \Vertex(38.79,-21.21){1}
\ArrowArc(60,0)(30,225,315)      \Vertex(81.21,-21.21){1}
\ArrowArc(60,0)(30,315,0)        \Vertex(90,0){1}
\PhotonArc(60,-42.42)(30,45,135){4}{3.5}
\Text(60,-7)[b]{$Z$,$\gamma$}
\Photon(90,0)(120,0){4}{2.5}
\Text(120,7)[br]{$W^+_\nu$}
\Text(60,-38)[t]{(o)}
\end{picture}
\caption{${\cal O}(N_f\alpha^2)$ Feynman diagrams contributing to
         the $W$ boson self energy.
\label{fig:WSelfe}
         }
\end{center}
\end{figure}

The Feynman diagrams contributing to the $W$-boson self-energy at
${\cal O}(N_f\alpha^2)$ are shown in Fig.\ref{fig:WSelfe}.
Internal lines labeled $Z$,$\gamma$ mean that all allowable
combinations must be included.

For the present case
of zero external momentum, $q=0$, the vector boson self-energies,
$\Pi_{\mu\nu}(q^2)$, can only take the form
\[
\Pi_{\mu\nu}(0)=\delta_{\mu\nu}F
\]
where $F$ is a function of the internal masses only and may be obtained
from the tensor integral representation of $\Pi_{\mu\nu}(0)$ by means of
the projection operator, $\delta_{\mu\nu}/n$. Thus
\begin{equation}
F=\left(\frac{\delta_{\mu\nu}}{n}\right)\Pi_{\mu\nu}(0).
\end{equation}
The resulting scalar integral can always be written in terms of the
master integral,
$I_0(j,k,l,m,n,M^2)$, of eq.(\ref{eq:MasterIntexpr}) and the results
are exact for all $n$.

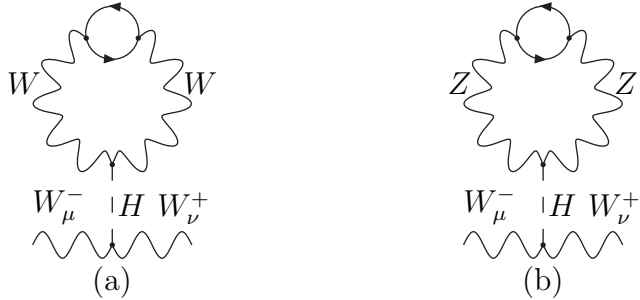
\begin{figure}
\begin{center}
\begin{picture}(65,100)(0,0)
\Photon(0,0)(30,0){5}{2}    \Vertex(30,0){1}
\Photon(30,0)(60,0){-5}{2}
\Text(0,8)[bl]{$W^-_\mu$}   \Text(67,8)[br]{$W^+_\nu$}
\PhotonArc(30,55.15)(24.85,-90,66.78){-5}{5}   \Vertex(20.20,77.99){1}
\PhotonArc(30,55.15)(24.85,113.22,270){5}{5} \Vertex(39.80,77.99){1}
\Text( 3,57)[br]{$W$}
\Text(57,57)[bl]{$W$}
\ArrowArc(30,80)(10,-11.61,181.61)
\ArrowArc(30,80)(10,181.61,-11.61)
\DashLine(30,30.30)(30,0){6} \Vertex(30,30.30){1}
\Text(32,15.15)[l]{$H$}
\Text(30,-8)[t]{(a)}
\end{picture}
\qquad\qquad\qquad\qquad
\begin{picture}(65,100)(0,0)
\Photon(0,0)(30,0){5}{2}    \Vertex(30,0){1}
\Photon(30,0)(60,0){-5}{2}
\Text(0,8)[bl]{$W^-_\mu$}   \Text(67,8)[br]{$W^+_\nu$}
\PhotonArc(30,55.15)(24.85,-90,66.78){-5}{5}   \Vertex(20.20,77.99){1}
\PhotonArc(30,55.15)(24.85,113.22,270){5}{5} \Vertex(39.80,77.99){1}
\Text( 3,57)[br]{$Z$}
\Text(57,57)[bl]{$Z$}
\ArrowArc(30,80)(10,-11.61,181.61)
\ArrowArc(30,80)(10,181.61,-11.61)
\DashLine(30,30.30)(30,0){6} \Vertex(30,30.30){1}
\Text(32,15.15)[l]{$H$}
\Text(30,-8)[t]{(b)}
\end{picture}
\vglue 5mm
\caption{${\cal O}(N_f\alpha^2)$ tadpole diagrams contributing to
         the $W$ boson self energy.
\label{fig:WTadpole}
         }
\end{center}
\end{figure}

Tadpole diagrams that can appear in the $W$ self-energy at
${\cal O}(N_f\alpha^2)$ are shown in Fig.\ref{fig:WTadpole}.
At this order the
individual diagrams are gauge-invariant and they are the only
contributions $\propto M_H^{-2}$ where $M_H$ is the Higgs mass.
In some renormalization schemes it is possible to eliminate them by a
suitable choice of the tadpole counterterm, $\delta\beta$, however in a
strictly $\overline{{\rm MS}}$ calculation this is not an option.
The finite parts of tadpole diagrams are not expected to enter strongly
into physical results since they represent a universal shift in the
Higgs vacuum expectation value and thus they have not been
included with the other corrections.

Since the Higgs mass is now known to satisfy, $M_H>M_W$, one may define,
for notational convenience,
$c_h^2=1-M_W^2/M_H^2$ and $s_h^2=1-c_h^2$ in analogy with $c_\theta$ and
$s_\theta$.

Fig.\ref{fig:WSelfe}(c), along with associated counterterm diagrams,
is the only topology in which the physical Higgs particle occurs.
In the $\overline{{\rm MS}}$ renormalization scheme
the 1-loop counterterm insertions in 1-loop diagrams
containing the Higgs vanish when taken together.
Fig.\ref{fig:WSelfe}(c) therefore accounts for the full $M_H$ dependence
in $\Delta r^{(2)}$. In order to obtain the ${\cal O}(N_f\alpha^2)$
$W$ boson mass counterterm, $\delta M_W^{2(2)}$, the $W$ boson
self-energy, $\Pi_{WW}(q^2)$ needs to be evaluated
at $q^2=-M_W^2$. In contrast the other ${\cal O}(N_f\alpha^2)$
counterterms that occur in the calculation can be gotten from Feynman
diagrams evaluated at $q^2=0$. In principle $\Pi_{WW}^{(2)}(-M_W^2)$
could be obtained in its entirety by the methods described in
ref.s\cite{WeigSchaBohm,ScharfTausk} however in the $\overline{{\rm MS}}$
renormalization scheme only its divergent pieces are needed and these
are considerably easier to extract and yield much more manageable
expressions.
Suffice it to say that they contain both local pieces, proportional
to polynomials in $q^2$ and non-local pieces proportional to
$\ln q^2$, $\ln(q^2+M_W^2)$ etc. When combined with 1-loop diagrams in
which 1-loop counterterms have been inserted the non-local divergences
must cancel so that the remaining divergences
can be removed by purely local
${\cal O}(N_f\alpha^2)$ counterterms. This cancellation occurs between
several diagrams and provides a stringent check of relative signs and
combinatoric factors.

The diagrams involving an internal photon require additional care due
to the fact that the integrals encounter singularities in certain
limits of interest for the external momenta.  Such diagrams
were evaluated by introducing a small mass term for the photon
and then taking the limit in an appropriate order.  In the
diagrams of Fig.\ref{fig:WSelfe}(a)\&(d), $q^2=-M_W^2$ is a branch
point and the integral blows up when evaluated na\"\i vely. In that
case the photon mass is set to zero only
after taking the limit $q\rightarrow -M_W^2$.
Further checks on the procedure were obtained by evaluating the diagrams
in various regions of momentum space with and without the photon mass
term; for example,
the $q\rightarrow0$ limits were checked against the same diagrams
evaluated by setting $q=0$ at the outset.

At ${\cal O}(N_f\alpha^2)$ the 2-point $W$ counterterm is
\begin{multline*}
\begin{picture}(32,5)(0,0)
\Photon(0,0)(68,0){4}{6}
\LongArrow(0,-8)(17,-8)
\Text(17,-12)[t]{$q$}
\Text(0,19)[tl]{$W^\pm_\mu$}
\Text(68,19)[tr]{$W^\mp_\nu$}
\SetWidth{1.0}
\Line(26,8)(40,-6)
\Line(24,6)(38,-8)
\Line(24,-6)(38,8)
\Line(26,-8)(40,6)
\end{picture}
\qquad\qquad
=-\left(q^2+M_W^2\right)\delta Z_W^{(2)}\delta_{\mu\nu}
 -\delta Z_W^{(2)}q_\mu q_\nu
 -\delta M_W^{2(2)}\delta_{\mu\nu}\\
 -\left(\delta Z_W^{(1f)}\delta M_W^{2(1b)}
       +\delta Z_W^{(1b)}\delta M_W^{2(1f)}\right)\delta_{\mu\nu}.
\end{multline*}

The $W$ boson self-energy enters $\Delta r$ via
the relation $\Delta r^{(2)}_{{\rm SE}}=\Pi^{(2)}_{WW}(0)/M_W^2$.
For one complete generation of massless fermions the
${\cal O}(N_f\alpha^2)$ diagrams containing the physical Higgs,
Fig.\ref{fig:WSelfe}(c), gives
\begin{multline}
\Delta r^{(2)}_{{\rm SE}_H}=
-\left(\frac{g^2}{16\pi^2}\right)^2
\bigg\{\frac{(20+s_h^2+2\pi^2 s_h^2)}{8s_h^2}
    -\frac{(4+s_h^2)}{2s_h^2}\ln\frac{M_W^2}{\mu^{\prime2}}\\
    -\frac{\ln c_h^2}{s_h^4}
          \left(\ln\frac{M_W^2}{\mu^{\prime2}}
               +\ln\frac{M_H^2}{\mu^{\prime2}}-\frac{5}{2}\right)
    +\ln^2\frac{M_W^2}{\mu^{\prime2}}
    \bigg\}
\label{eq:DeltarSEH}
\end{multline}

After combining the ${\cal O}(N_f\alpha^2)$ diagrams of
Fig.\ref{fig:WSelfe} with
1-loop diagrams that contain 1-loop counter\-term insertions the result
is
\begin{eqnarray}
\Delta r^{(2)}_{{\rm SE}_W}&=&
 \left(\frac{g^2}{16\pi^2}\right)^2\bigg\{
 \frac{(1158-3496s_\theta^2+2803s_\theta^4-480s_\theta^6)}{72c_\theta^4}
+\frac{(4-8s_\theta^2-s_\theta^4)}{12c_\theta^4}\pi^2\nonumber\\
 & & \qquad\qquad
+\frac{(294-529s_\theta^2)}{18s_\theta^2}\ln c_\theta^2
-\frac{(12-23s_\theta^2)}{3s_\theta^2}
      \left(\ln c_\theta^2+2\ln \frac{M_Z^2}{\mu^{\prime2}}\right)
      \ln c_\theta^2
\label{eq:DeltarSEW}\\
 & & \qquad\qquad
-\frac{(50-156s_\theta^2+117s_\theta^4-16s_\theta^6)}{6c_\theta^4}
      \ln\frac{M_Z^2}{\mu^{\prime2}}
+\frac{(3-6s_\theta^2-2s_\theta^4)}{3c_\theta^4}
      \ln^2\frac{M_Z^2}{\mu^{\prime2}}
\bigg\}\nonumber
\end{eqnarray}
As noted above the divergences that remained have been shown to be purely
local and have been removed in a manner consistent with
$\overline{{\rm MS}}$ renormalization.

\subsection{Vertex Corrections}

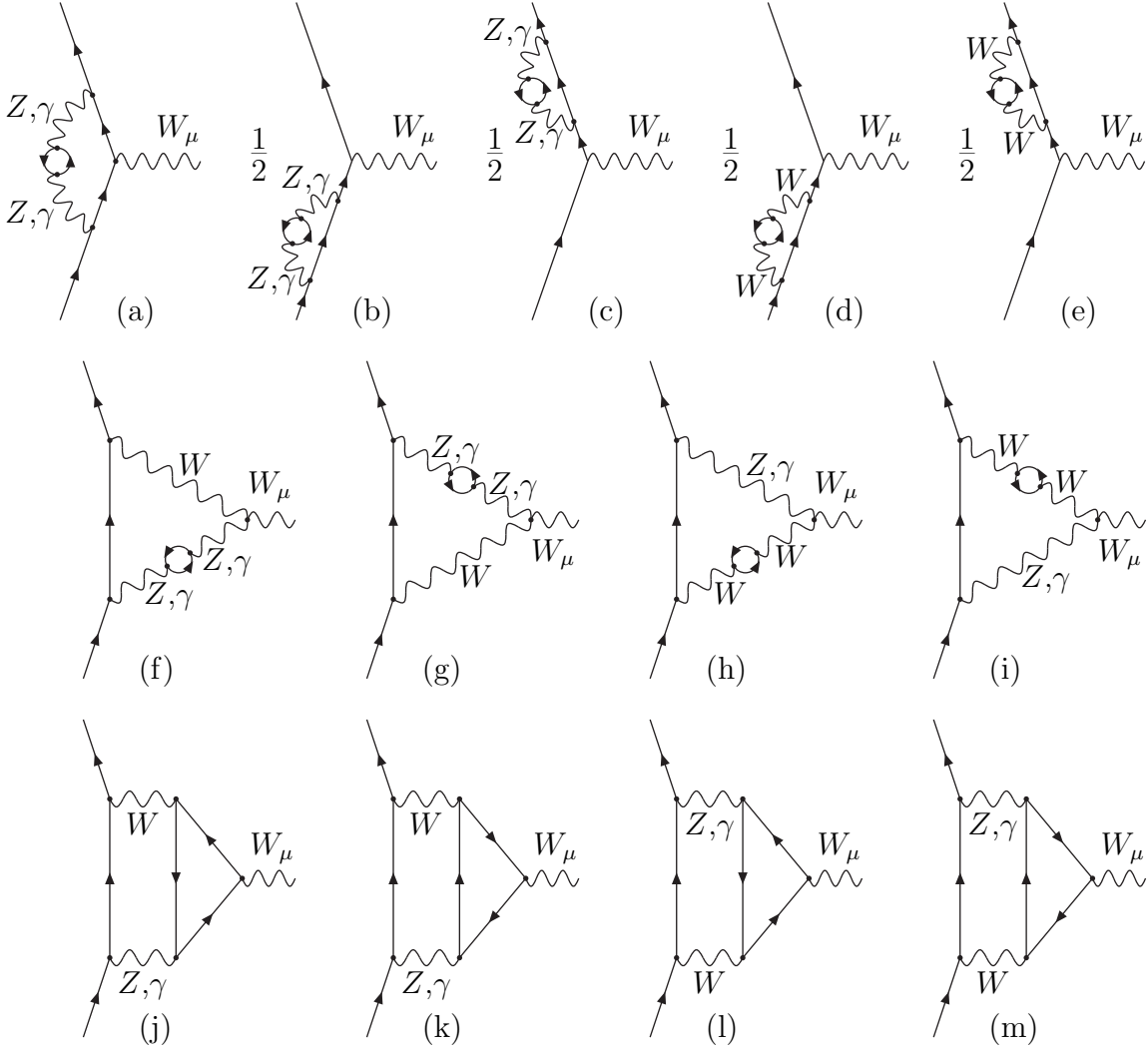
\begin{figure}
\begin{center}
\begin{picture}(62,120)(0,0)
\ArrowLine(0,0)(12.25,35)   \Vertex(12.25,35){1}
\ArrowLine(12.25,35)(21,60) \Vertex(21,60){1}
\ArrowLine(21,60)(12.25,85) \Vertex(12.25,85){1}
\ArrowLine(12.25,85)(0,120)
\PhotonArc(19.25,65)(21.08,109.29,180){3}{3.5}
\Text(-2,80)[r]{$Z$,$\gamma$}
\ArrowArc(-1.08,60)(5,90,-90)  \Vertex(-1.08,65){1}
\ArrowArc(-1.08,60)(5,-90,90)  \Vertex(-1.08,55){1}
\PhotonArc(19.25,55)(21.08,180,250.71){3}{3.5}
\Text(-2,40)[r]{$Z$,$\gamma$}
\Photon(21,60)(53,60){3}{4}
\Text(53,66)[br]{$W_\mu$}
\Text(21,-3)[bl]{(a)}
\end{picture}
\qquad
\begin{picture}(62,120)(0,0)
\ArrowLine(0,0)(5.25,15)       \Vertex(5.25,15){1}
\ArrowLine(5.25,15)(15.75,45)  \Vertex(15.75,45){1}
\ArrowLine(15.75,45)(21,60)
\ArrowLine(21,60)(0,120)
\PhotonArc(8.35,25.28)(10.89,160.71,250.71){3}{2.5}
\Text(0,15)[r]{$Z$,$\gamma$}
\ArrowArc(0.22,33.60)(5,70.71,250.71)  \Vertex(1.87,38.32){1}
\ArrowArc(0.22,33.60)(5,250.71,70.71)  \Vertex(-1.43,28.88){1}
\PhotonArc(12.15,34.72)(10.89,70.71,160.71){3}{2.5}
\Text(13,53)[r]{$Z$,$\gamma$}
\Photon(21,60)(53,60){3}{4}
\Text(53,66)[br]{$W_\mu$}
\Text(-17,65)[bl]{1} \Text(-17,60)[bl]{--} \Text(-17,53)[bl]{2}
\Text(21,-3)[bl]{(b)}
\end{picture}
\qquad
\begin{picture}(62,120)(0,0)
\ArrowLine(0,0)(21,60)
\ArrowLine(21,60)(15.75,75)     \Vertex(15.75,75){1}
\ArrowLine(15.75,75)(5.25,105)  \Vertex(5.25,105){1}
\ArrowLine(5.25,105)(0,120)
\PhotonArc(8.35,94.72)(10.89,109.29,199.29){3}{2.5}
\Text(0,109)[r]{$Z$,$\gamma$}
\PhotonArc(12.15,85.28)(10.89,199.29,289.29){3}{2.5}
\Text(12,70)[r]{$Z$,$\gamma$}
\ArrowArc(0.22,86.40)(5,109.29,289.29)  \Vertex(1.87,81.68){1}
\ArrowArc(0.22,86.40)(5,289.29,109.29)  \Vertex(-1.43,91.12){1}
\Photon(21,60)(53,60){3}{4}
\Text(53,66)[br]{$W_\mu$}
\Text(-17,65)[bl]{1} \Text(-17,60)[bl]{--} \Text(-17,53)[bl]{2}
\Text(21,-3)[bl]{(c)}
\end{picture}
\qquad
\begin{picture}(62,120)(0,0)
\ArrowLine(0,0)(5.25,15)       \Vertex(5.25,15){1}
\ArrowLine(5.25,15)(15.75,45)  \Vertex(15.75,45){1}
\ArrowLine(15.75,45)(21,60)
\ArrowLine(21,60)(0,120)
\PhotonArc(8.35,25.28)(10.89,160.71,250.71){3}{2.5}
\Text(0,13)[r]{$W$}
\ArrowArc(0.22,33.60)(5,70.71,250.71)  \Vertex(1.87,38.32){1}
\ArrowArc(0.22,33.60)(5,250.71,70.71)  \Vertex(-1.43,28.88){1}
\PhotonArc(12.15,34.72)(10.89,70.71,160.71){3}{2.5}
\Text(15,53)[r]{$W$}
\Photon(21,60)(53,60){3}{4}
\Text(51,66)[br]{$W_\mu$}
\Text(-17,65)[bl]{1} \Text(-17,60)[bl]{--} \Text(-17,53)[bl]{2}
\Text(21,-3)[bl]{(d)}
\end{picture}
\qquad
\begin{picture}(62,120)(0,0)
\ArrowLine(0,0)(21,60)
\ArrowLine(21,60)(15.75,75)     \Vertex(15.75,75){1}
\ArrowLine(15.75,75)(5.25,105)  \Vertex(5.25,105){1}
\ArrowLine(5.25,105)(0,120)
\PhotonArc(8.35,94.72)(10.89,109.29,199.29){3}{2.5}
\Text(-2,103)[r]{$W$}
\PhotonArc(12.15,85.28)(10.89,199.29,289.29){3}{2.5}
\Text(12,67)[r]{$W$}
\ArrowArc(0.22,86.40)(5,109.29,289.29)  \Vertex(1.87,81.68){1}
\ArrowArc(0.22,86.40)(5,289.29,109.29)  \Vertex(-1.43,91.12){1}
\Photon(21,60)(53,60){3}{4}
\Text(53,66)[br]{$W_\mu$}
\Text(-17,65)[bl]{1} \Text(-17,60)[bl]{--} \Text(-17,53)[bl]{2}
\Text(21,-3)[bl]{(e)}
\end{picture}
\\\vglue 15pt
\begin{picture}(80,120)(0,0)
\ArrowLine(0,0)(10,30)      \Vertex(10,30){1}
\ArrowLine(10,30)(10,90)    \Vertex(10,90){1}
\ArrowLine(10,90)(0,120)
\Photon(10,30)(31.65,42.5){-3}{2.5}
\Text(23,36.5)[tl]{$Z$,$\gamma$}
\Photon(40.31,47.5)(61.96,60){-3}{2.5}    \Vertex(61.96,60){1}
\Text(45,49.5)[tl]{$Z$,$\gamma$}
\Photon(61.96,60)(10,90){-3}{5.5}
\Text(36,77)[bl]{$W$}
\Photon(61.96,60)(80,60){3}{2}
\Text(79,67)[br]{$W_\mu$}
\ArrowArc(35.98,45)(5,-120,60)    \Vertex(40.31,47.5){1}
\ArrowArc(35.98,45)(5,60,-120)    \Vertex(31.65,42.5){1}
\Text(21,-3)[bl]{(f)}
\end{picture}
\qquad
\begin{picture}(80,120)(0,0)
\ArrowLine(0,0)(10,30)      \Vertex(10,30){1}
\ArrowLine(10,30)(10,90)    \Vertex(10,90){1}
\ArrowLine(10,90)(0,120)
\Photon(10,90)(31.65,77.5){3}{2.5}
\Text(24,81.5)[bl]{$Z$,$\gamma$}
\Photon(40.31,72.5)(61.96,60){3}{2.5}    \Vertex(61.96,60){1}
\Text(46,67.5)[bl]{$Z$,$\gamma$}
\Photon(61.96,60)(10,30){3}{5.5}
\Text(35,43)[tl]{$W$}
\Photon(61.96,60)(80,60){3}{2}
\Text(79,53)[tr]{$W_\mu$}
\ArrowArc(35.98,75)(5,120,300)    \Vertex(40.31,72.5){1}
\ArrowArc(35.98,75)(5,300,120)    \Vertex(31.65,77.5){1}
\Text(21,-3)[bl]{(g)}
\end{picture}
\qquad
\begin{picture}(80,120)(0,0)
\ArrowLine(0,0)(10,30)      \Vertex(10,30){1}
\ArrowLine(10,30)(10,90)    \Vertex(10,90){1}
\ArrowLine(10,90)(0,120)
\Photon(10,30)(31.65,42.5){-3}{2.5}
\Text(24,36.5)[tl]{$W$}
\Photon(40.31,47.5)(61.96,60){-3}{2.5}    \Vertex(61.96,60){1}
\Text(47,50.5)[tl]{$W$}
\Photon(61.96,60)(10,90){-3}{5.5}
\Text(36,75)[bl]{$Z$,$\gamma$}
\Photon(61.96,60)(80,60){3}{2}
\Text(79,67)[br]{$W_\mu$}
\ArrowArc(35.98,45)(5,-120,60)    \Vertex(40.31,47.5){1}
\ArrowArc(35.98,45)(5,60,-120)    \Vertex(31.65,42.5){1}
\Text(21,-3)[bl]{(h)}
\end{picture}
\qquad
\begin{picture}(80,120)(0,0)
\ArrowLine(0,0)(10,30)      \Vertex(10,30){1}
\ArrowLine(10,30)(10,90)    \Vertex(10,90){1}
\ArrowLine(10,90)(0,120)
\Photon(10,90)(31.65,77.5){3}{2.5}
\Text(24,84.5)[bl]{$W$}
\Photon(40.31,72.5)(61.96,60){3}{2.5}    \Vertex(61.96,60){1}
\Text(46,70.5)[bl]{$W$}
\Photon(61.96,60)(10,30){3}{5.5}
\Text(33,43)[tl]{$Z$,$\gamma$}
\Photon(61.96,60)(80,60){3}{2}
\Text(79,53)[tr]{$W_\mu$}
\ArrowArc(35.98,75)(5,120,300)    \Vertex(40.31,72.5){1}
\ArrowArc(35.98,75)(5,300,120)    \Vertex(31.65,77.5){1}
\Text(21,-3)[bl]{(i)}
\end{picture}
\\\vglue 15pt
\begin{picture}(80,120)(0,0)
\ArrowLine(0,0)(10,30)      \Vertex(10,30){1}
\ArrowLine(10,30)(10,90)    \Vertex(10,90){1}
\ArrowLine(10,90)(0,120)
\Photon(10,30)(35,30){-3}{2.5}
\Text(22.5,25)[t]{$Z$,$\gamma$}
\Photon(10,90)(35,90){-3}{2.5}
\Text(22.5,85)[t]{$W$}
\ArrowLine(35,90)(35,30)       \Vertex(35,90){1}
\ArrowLine(35,30)(60,60)       \Vertex(35,30){1}
\ArrowLine(60,60)(35,90)       \Vertex(60,60){1}
\Photon(60,60)(80,60){-3}{2.5}
\Text(80,66)[br]{$W_\mu$}
\Text(21,-3)[bl]{(j)}
\end{picture}
\qquad
\begin{picture}(80,120)(0,0)
\ArrowLine(0,0)(10,30)      \Vertex(10,30){1}
\ArrowLine(10,30)(10,90)    \Vertex(10,90){1}
\ArrowLine(10,90)(0,120)
\Photon(10,30)(35,30){-3}{2.5}
\Text(22.5,25)[t]{$Z$,$\gamma$}
\Photon(10,90)(35,90){-3}{2.5}
\Text(22.5,85)[t]{$W$}
\ArrowLine(35,30)(35,90)       \Vertex(35,90){1}
\ArrowLine(60,60)(35,30)       \Vertex(35,30){1}
\ArrowLine(35,90)(60,60)       \Vertex(60,60){1}
\Photon(60,60)(80,60){-3}{2.5}
\Text(80,66)[br]{$W_\mu$}
\Text(21,-3)[bl]{(k)}
\end{picture}
\qquad
\begin{picture}(80,120)(0,0)
\ArrowLine(0,0)(10,30)      \Vertex(10,30){1}
\ArrowLine(10,30)(10,90)    \Vertex(10,90){1}
\ArrowLine(10,90)(0,120)
\Photon(10,30)(35,30){-3}{2.5}
\Text(22.5,25)[t]{$W$}
\Photon(10,90)(35,90){-3}{2.5}
\Text(22.5,85)[t]{$Z$,$\gamma$}
\ArrowLine(35,90)(35,30)       \Vertex(35,90){1}
\ArrowLine(35,30)(60,60)       \Vertex(35,30){1}
\ArrowLine(60,60)(35,90)       \Vertex(60,60){1}
\Photon(60,60)(80,60){-3}{2.5}
\Text(80,66)[br]{$W_\mu$}
\Text(21,-3)[bl]{(l)}
\end{picture}
\qquad
\begin{picture}(80,120)(0,0)
\ArrowLine(0,0)(10,30)      \Vertex(10,30){1}
\ArrowLine(10,30)(10,90)    \Vertex(10,90){1}
\ArrowLine(10,90)(0,120)
\Photon(10,30)(35,30){-3}{2.5}
\Text(22.5,25)[t]{$W$}
\Photon(10,90)(35,90){-3}{2.5}
\Text(22.5,85)[t]{$Z$,$\gamma$}
\ArrowLine(35,30)(35,90)       \Vertex(35,90){1}
\ArrowLine(60,60)(35,30)       \Vertex(35,30){1}
\ArrowLine(35,90)(60,60)       \Vertex(60,60){1}
\Photon(60,60)(80,60){-3}{2.5}
\Text(80,66)[br]{$W_\mu$}
\Text(21,-3)[bl]{(m)}
\end{picture}
\end{center}
\caption{${\cal O}(N_f\alpha^2)$ vertex and external leg corrections
         contributing to muon decay.
\label{fig:VertexDiags}
        }
\end{figure}

The ${\cal O}(N_f\alpha^2)$ vertex diagrams and external leg corrections
contributing to $\Delta r^{(2)}$ are shown in Fig.\ref{fig:VertexDiags}.

The diagrams of Fig.\ref{fig:VertexDiags}
containing only virtual photons are IR
divergent and must be separated into UV finite, IR divergent QED
corrections that are already included by $\Delta q$ in the
extraction of $G_F$ \cite{muonprl} .
Sirlin \cite{Sirlin80,Sirlin78,Sirlin84} has described a strategy
that, starting from the full electroweak theory, makes the separation
of contributions to $\Delta q$ and $\Delta r$ automatic at least up
to ${\cal O}(\alpha m_\mu^2/M_W^2)$. In diagrams exhibiting infrared
(IR) divergences, the photon propagator is replaced by
\begin{equation}
\frac{1}{k^2}\longrightarrow\left\{\frac{1}{k^2}-\frac{1}{k^2
                                  +\Lambda^2}\right\}
                           +\frac{1}{k^2+\Lambda^2}.
\label{eq:photonsplit}
\end{equation}
where it is generally convenient to take $\Lambda=M_W$.
The term in curly brackets is simply the original photon propagator with
a Pauli-Villars regulator. It has the same IR behaviour
and gives contributions that are identical to those of Fermi
theory up to ${\cal O}(\alpha m_\mu^2/M_W^2)$ and thus are contained
in $\Delta q$. The second term in (\ref{eq:photonsplit}) gives
contributions that retain the original UV behaviour but are free from
IR singularities and therefore belong in $\Delta r$.
The UV divergent, IR finite weak corrections that are included
in $\Delta r^{(2)}$ are, as already pointed out in ref.\cite{muon1ppm},
independent of the separation mass, $\Lambda$,
because of a cancellation against corresponding 1-loop
diagrams with 1-loop counterterm insertions.

For processes with massless external fermions the only relevant
vertex corrections are those involving vector bosons and these will
necessarily be purely vector and axial-vector in character. A general
vertex correction can then be represented as
\[
\begin{picture}(43,42)(0,0)
\ArrowLine(0,-36)(12,0)
\ArrowLine(12,0)(0,36)
\Photon(12,0)(40,0){4}{3}
\BCirc(12,0){8}
\end{picture}
\equiv V_\mu=i\gamma_\mu(V_L\gamma_L+V_R\gamma_R)
\]
\vglue 1cm\noindent%
where $V_L$ and $V_R$ are functions only of the internal masses. The tensor
integral representation of $V_\mu$ can easily be obtained by standard
techniques and from it the scalar integral representations of $V_L$ and
$V_R$ follow by means of projection operators. Thus
\begin{equation}
V_{L,R}=-\frac{i}{2n}\Tr\{V_\mu\gamma_\mu\gamma_{R,L}\}.
\label{eq:vertexproj}
\end{equation}
where $\Tr\{\gamma_\mu\gamma_\mu\}=4n$ is assumed.
This method for directly obtaining the scalar integral representation
of the vertex form factors is particularly convenient when computer
algebra is being employed. Once again the resulting scalar integrals can
be written in terms of the master integral, $I_0(j,k,l,m,n,M^2)$.
of eq.(\ref{eq:MasterIntexpr}).

The ${\cal O}(N_f\alpha^2)$ $Wff^\prime$ vertex counterterm is
given by
\begin{equation}
\begin{picture}(43,42)(0,0)
\ArrowLine(0,-36)(12,0)
\Text(5,-36)[bl]{$f$}
\ArrowLine(12,0)(0,36)
\Text(5,36)[tl]{$f^\prime$}
\Photon(12,0)(40,0){4}{3}
\Text(40,6)[br]{$W_\mu$}
\SetWidth{1.0}
\Line(6,8)(20,-6)
\Line(4,6)(18,-8)
\Line(4,-6)(18,8)
\Line(6,-8)(20,6)
\end{picture}
=i\frac{g}{\sqrt{2}}\gamma_\mu\gamma_L\left\{
           \frac{1}{2}\delta Z_W^{(2)}+\frac{\delta g^{(2)}}{g}
          +2\frac{\delta g^{(1f)}}{g}
                 \left(\frac{1}{2}\delta Z_W^{(1b)}
                      +\frac{\delta g^{(1b)}}{g}\right)
          -3\frac{\delta g^{(1f)}}{g}.\frac{\delta g^{(1b)}}{g}\right\}
\label{eq:VertexCT2}
\end{equation}
\vglue 23pt\noindent%
In ref.\cite{MaldeStuart1} it was shown, by considering electric
charge renormalization, that in any renormalization scheme the
${\cal O}(N_f\alpha^2)$ counterterms must satisfy the relation
\begin{align}
\frac{1}{2}\delta Z_W^{(2)}
&+\frac{\delta g^{(2)}}{g}
+2\frac{\delta g^{(1f)}}{g}
     \left(\frac{1}{2}\delta Z_W^{(1b)}+\frac{\delta g^{(1b)}}{g}\right)
-3\frac{\delta g^{(1b)}}{g}.\frac{\delta g^{(1f)}}{g}\notag\\
=&\left(\frac{g^2}{16\pi^2}\right)^2
 8\frac{(\pi M_W^2)^{n-4}}{n}
\Gamma(4-n)\Gamma\left(2-\frac{n}{2}\right)\Gamma\left(\frac{n}{2}\right)
\notag\\
&-3\frac{\delta g^{(1f)}}{g}
\left(\frac{g^2}{16\pi^2}\right)
(\pi M_W^2)^{-\epsilon}\Gamma(\epsilon)
+\frac{\delta M_W^{2(1f)}}{M_W^2}
\left(\frac{g^2}{16\pi^2}\right)
(\pi M_W^2)^{-\epsilon}\epsilon\Gamma(\epsilon)
+\mbox{finite}
\label{eq:VertexDiv}
\end{align}
from which expressions for the $\overline{{\rm MS}}$ counterterms
on the left hand side of Eq.(\ref{eq:VertexDiv}) are easily extracted
giving
\begin{equation}
\frac{1}{2}\delta Z_W^{(2)}
+\frac{\delta g^{(2)}}{g}
+2\frac{\delta g^{(1f)}}{g}
     \left(\frac{1}{2}\delta Z_W^{(1b)}+\frac{\delta g^{(1b)}}{g}\right)
-3\frac{\delta g^{(1b)}}{g}.\frac{\delta g^{(1f)}}{g}=
-\frac{1}{\epsilon^2}+\frac{5}{6\epsilon}.
\label{eq:VertexCT}
\end{equation}

When the weak parts of the Feynman diagrams of Fig.\ref{fig:VertexDiags},
obtained by means of eq.(\ref{eq:photonsplit}), are combined with
the 1-loop diagrams with ${\cal O}(N_f\alpha)$ counterterm insertions and the
${\cal O}(N_f\alpha^2)$ counterterms (\ref{eq:VertexCT})
the result is indeed finite. This provides not only a check of the
calculation performed here but also of overall renormalization
prescription as performed in ref.\cite{MaldeStuart1}. The vertex and
external leg corrections for a
complete generation of massless fermions contribute
\begin{multline}
\Delta r^{(2)}_{{\rm vertex}}=-\left(\frac{g^2}{16\pi^2}\right)^2\bigg\{
  \frac{2\pi^2}{3}+\frac{5}{6}(17-32s_\theta^2)
 +\frac{4(11-27s_\theta^2)}{3s_\theta^2}\ln c_\theta^2\\
 -\frac{16}{3}(2-3s_\theta^2)\ln\frac{M_Z^2}{\mu^{\prime2}}
 -\frac{(5-12s_\theta^2)}{s_\theta^2}
   \left(\ln c_\theta^2+2\ln\frac{M_Z^2}{\mu^{\prime2}}\right)
   \ln c_\theta^2
 +2\ln^2\frac{M_Z^2}{\mu^{\prime2}}
   \bigg\}
\label{eq:DeltarVertex}
\end{multline}

\subsection{Box Diagrams}
\label{sect:BoxDiagrams}

\begin{figure}
\begin{center}
\begin{picture}(80,120)(0,0)
\ArrowLine(0,0)(10,30)       \Vertex(10,30){1}
\ArrowLine(10,30)(10,90)     \Vertex(10,90){1}
\ArrowLine(10,90)(0,120)
\Text(7,15)[l]{$\mu^-$}     \Text(8,104)[l]{$\nu_\mu$}
\ArrowLine(70,90)(80,120)    \Vertex(70,90){1}
\ArrowLine(70,30)(70,90)     \Vertex(70,30){1}
\ArrowLine(80,60)(70,30)
\Text(78,46)[l]{$\bar\nu_e$} \Text(78,106)[l]{$e^-$}
\Photon(10,30)(33,30){-4}{2}    \Vertex(33,30){1}
\Photon(47,30)(70,30){4}{2}     \Vertex(47,30){1}
\ArrowArc(40,30)(7,180,0)
\ArrowArc(40,30)(7,0,180)
\Text(43,82)[t]{$W$}
\Photon(10,90)(70,90){4}{5.5}
\Text(21.5,37)[b]{$Z$,$\gamma$} \Text(58.5,37)[b]{$Z$}
\Text(40,0)[b]{(a)}
\end{picture}
\qquad
\begin{picture}(80,120)(0,0)
\ArrowLine(0,0)(10,30)       \Vertex(10,30){1}
\ArrowLine(10,30)(10,90)     \Vertex(10,90){1}
\ArrowLine(10,90)(0,120)
\Text(7,15)[l]{$\mu^-$}     \Text(8,104)[l]{$\nu_\mu$}
\ArrowLine(70,90)(80,120)    \Vertex(70,90){1}
\ArrowLine(70,30)(70,90)     \Vertex(70,30){1}
\ArrowLine(80,60)(70,30)
\Text(78,46)[l]{$\bar\nu_e$} \Text(78,106)[l]{$e^-$}
\Photon(10,90)(33,90){4}{2}     \Vertex(33,90){1}
\Photon(47,90)(70,90){-4}{2}    \Vertex(47,90){1}
\ArrowArc(40,90)(7,180,0)
\ArrowArc(40,90)(7,0,180)
\Text(43,37)[b]{$W$}
\Photon(10,30)(70,30){-4}{5.5}
\Text(21.5,82)[t]{$Z$} \Text(58.5,82)[t]{$Z$,$\gamma$}
\Text(40,0)[b]{(b)}
\end{picture}
\qquad
\begin{picture}(80,120)(0,0)
\ArrowLine(0,0)(10,30)       \Vertex(10,30){1}
\ArrowLine(10,30)(10,90)     \Vertex(10,90){1}
\ArrowLine(10,90)(0,120)
\Text(7,15)[l]{$\mu^-$}     \Text(8,104)[l]{$\nu_\mu$}
\ArrowLine(70,90)(80,120)    \Vertex(70,90){1}
\ArrowLine(70,30)(70,90)     \Vertex(70,30){1}
\ArrowLine(80,60)(70,30)
\Text(78,46)[l]{$\bar\nu_e$} \Text(78,106)[l]{$e^-$}
\Photon(10,30)(33,30){-4}{2}    \Vertex(33,30){1}
\Photon(47,30)(70,30){4}{2}     \Vertex(47,30){1}
\ArrowArc(40,30)(7,180,0)
\ArrowArc(40,30)(7,0,180)
\Text(43,83)[t]{$Z$}
\Photon(10,90)(70,90){4}{5.5}
\Text(21.5,37)[b]{$W$} \Text(58.5,37)[b]{$W$}
\Text(40,0)[b]{(c)}
\end{picture}
\qquad
\begin{picture}(80,120)(0,0)
\ArrowLine(0,0)(10,30)       \Vertex(10,30){1}
\ArrowLine(10,30)(10,90)     \Vertex(10,90){1}
\ArrowLine(10,90)(0,120)
\Text(7,15)[l]{$\mu^-$}     \Text(8,104)[l]{$\nu_\mu$}
\ArrowLine(70,90)(80,120)    \Vertex(70,90){1}
\ArrowLine(70,30)(70,90)     \Vertex(70,30){1}
\ArrowLine(80,60)(70,30)
\Text(78,46)[l]{$\bar\nu_e$} \Text(78,106)[l]{$e^-$}
\Photon(10,90)(33,90){4}{2}     \Vertex(33,90){1}
\Photon(47,90)(70,90){-4}{2}    \Vertex(47,90){1}
\ArrowArc(40,90)(7,180,0)
\ArrowArc(40,90)(7,0,180)
\Text(43,37)[b]{$Z$}
\Photon(10,30)(70,30){-4}{5.5}
\Text(21.5,82)[t]{$W$} \Text(58.5,82)[t]{$W$}
\Text(40,0)[b]{(d)}
\end{picture}
\\\vglue 15pt
\caption{${\cal O}(N_f\alpha^2)$ box diagrams contributing
         to muon decay. Other box diagrams that can be constructed
         vanish.
\label{fig:BoxDiagrams}
         }
\end{center}
\end{figure}
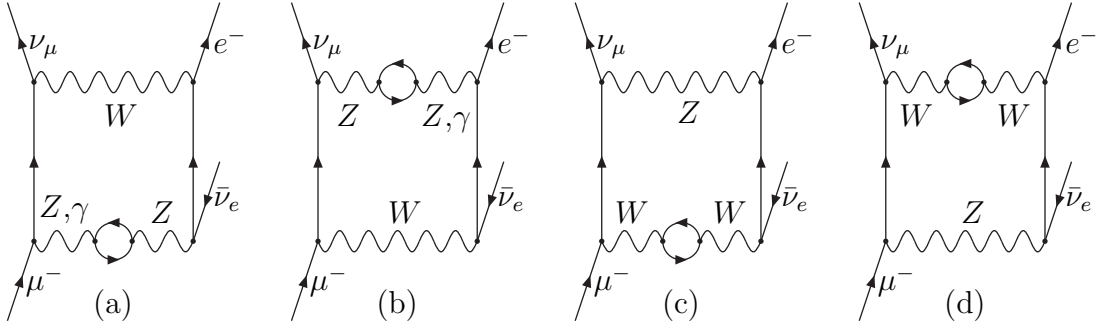

The ${\cal O}(N_f\alpha^2)$ box diagrams contributing to muon decay are
shown in Fig.\ref{fig:BoxDiagrams}. Other box diagrams can
be constructed in which the internal bosons cross but these all vanish
identically because they are proportional to
products of the left-handed with right-handed couplings of the $W$ for
which the latter is zero. The only IR divergent box diagram, that
contains an internal photon, vanishes in this way and so the
procedure for separating QED and weak corrections
(\ref{eq:photonsplit}) does not need to be invoked.
All diagrams are only logarithmically divergent with simple poles at
$n=4$.

The box diagrams containing a virtual photon and a counterterm insertion
on the $W$ propagator can also, in principle, produce IR divergent
contributions that would require the separation strategy
(\ref{eq:photonsplit}) to be invoked but
\begin{equation}
\begin{picture}(40,60)(0,0)
\ArrowLine(0,0)(5,15)
\ArrowLine(5,15)(5,45)
\ArrowLine(5,45)(0,60)
\ArrowLine(40,60)(35,45)
\ArrowLine(35,45)(35,15)     \Vertex(35,15){1}
\ArrowLine(35,15)(40,30)
\Photon(5,15)(35,15){-2}{5.5}
\Text(21.5,41)[t]{${\scriptstyle W}$}
\Photon(5,45)(35,45){2}{5.5}
\Text(21.5,11)[t]{${\scriptstyle \gamma}$}
\SetWidth{1}
\Line(1.5,41.5)(8.5,48.5)
\Line(1.5,48.5)(8.5,41.5)
\end{picture}
\quad\raisebox{30pt}{+}\quad
\begin{picture}(40,60)(0,0)
\ArrowLine(0,0)(5,15)        \Vertex(5,15){1}
\ArrowLine(5,15)(5,45)       \Vertex(5,45){1}
\ArrowLine(5,45)(0,60)
\ArrowLine(40,60)(35,45)     \Vertex(35,45){1}
\ArrowLine(35,45)(35,15)     \Vertex(35,15){1}
\ArrowLine(35,15)(40,30)
\Photon(5,15)(35,15){-2}{5.5}
\Text(11.5,41)[t]{${\scriptstyle W}$} \Text(28.5,41)[t]{${\scriptstyle W}$}
\Photon(5,45)(35,45){2}{5.5}
\Text(21.5,11)[t]{${\scriptstyle \gamma}$}
\SetWidth{1}
\Line(16.5,41.5)(23.5,48.5)
\Line(16.5,48.5)(23.5,41.5)
\end{picture}
\quad\raisebox{30pt}{+}\quad
\begin{picture}(40,60)(0,0)
\ArrowLine(0,0)(5,15)        \Vertex(5,15){1}
\ArrowLine(5,15)(5,45)       \Vertex(5,45){1}
\ArrowLine(5,45)(0,60)
\ArrowLine(40,60)(35,45)     \Vertex(35,45){1}
\ArrowLine(35,45)(35,15)     \Vertex(35,15){1}
\ArrowLine(35,15)(40,30)
\Photon(5,15)(35,15){-2}{5.5}
\Text(21.5,41)[t]{${\scriptstyle W}$}
\Photon(5,45)(35,45){2}{5.5}
\Text(21.5,11)[t]{${\scriptstyle \gamma}$}
\SetWidth{1}
\Line(31.5,41.5)(38.5,48.5)
\Line(31.5,48.5)(38.5,41.5)
\end{picture}
\quad\raisebox{30pt}%
{$=\left(2\frac{{\displaystyle\delta g^{(1f)}}}{{\displaystyle g}}
         -\frac{{\displaystyle\delta M_W^{2(1f)}}}{{\displaystyle M_W^2}}
   \right)$}\quad
\begin{picture}(40,60)(0,0)
\ArrowLine(0,0)(5,15)        \Vertex(5,15){1}
\ArrowLine(5,15)(5,45)       \Vertex(5,45){1}
\ArrowLine(5,45)(0,60)
\ArrowLine(40,60)(35,45)     \Vertex(35,45){1}
\ArrowLine(35,45)(35,15)     \Vertex(35,15){1}
\ArrowLine(35,15)(40,30)
\Photon(5,15)(35,15){-2}{5.5}
\Text(21.5,41)[t]{${\scriptstyle W}$}
\Photon(5,45)(35,45){2}{5.5}
\Text(21.5,11)[t]{${\scriptstyle \gamma}$}
\end{picture}
\label{eq:boxCTS}
\end{equation}
and in the $\overline{{\rm MS}}$ renormalization scheme the
combination of counterterms on the right hand side of
Eq.(\ref{eq:boxCTS}) vanishes.

For 4-fermion processes one-loop box diagrams are finite but at
${\cal O}(N_f\alpha^2)$ they develop logarithmic divergences.
A useful set of identities for calculating one-loop box diagrams appears
in ref.\cite{SirlinBox}. They are, however, valid only for $n=4$
because of their intended use at one-loop.
The relations involve products of strings of three $\gamma$ matrices
and care must be taken in applying and generalizing identities like
\begin{equation}
\gamma_\rho\gamma_\alpha\gamma_\sigma=
 \delta_{\rho\alpha}\gamma_\sigma
-\delta_{\rho\sigma}\gamma_\alpha
+\delta_{\alpha\sigma}\gamma_\rho
-\epsilon_{\rho\alpha\sigma\beta}\gamma_\beta\gamma_5.
\end{equation}
It should be noted that at ${\cal O}(N_f\alpha^2)$ the box diagrams
contain only simple poles at $n=4$ and are rendered finite by
${\cal O}(\alpha)$ counterterms inserted into finite 1-loop diagrams.
The details of the convention adopted in treating terms proportional
to $(n-4)$ cancel and are therefore irrelevant for the present
purposes. The only requirement is that a consistent convention be
adopted. For definiteness we take
\vspace{0.5cm}

\noindent\parbox[t][1cm][b]{8cm}{
\begin{eqnarray*}
{[}\gamma_\rho\gamma_\mu\gamma_\sigma\gamma_{L,R}{]}_1
{[}\gamma_\rho\gamma_\nu\gamma_\sigma\gamma_{L,R}{]}_2
&=&4\delta_{\mu\nu}{[}\gamma_\alpha\gamma_{L,R}{]}_1
                       {[}\gamma_\alpha\gamma_{L,R}{]}_2
                 +(n-4){[}\gamma_\mu\gamma_{L,R}{]}_1
                       {[}\gamma_\nu\gamma_{L,R}{]}_2
\end{eqnarray*}\hfill} \hfill\parbox[t][1cm][c]{1cm}
{\hfill \begin{eqnarray}\label{eq:firstBoxid}\end{eqnarray}}
\noindent\parbox[t][1cm][b]{8cm}{
\begin{eqnarray*}
{[}\gamma_\rho\gamma_\mu\gamma_\sigma\gamma_{L,R}{]}_1
{[}\gamma_\rho\gamma_\nu\gamma_\sigma\gamma_{R,L}{]}_2
                   &=&4{[}\gamma_\nu\gamma_{L,R}{]}_1
                       {[}\gamma_\mu\gamma_{R,L}{]}_2
                 +(n-4){[}\gamma_\mu\gamma_{L,R}{]}_1
                       {[}\gamma_\nu\gamma_{R,L}{]}_2
\end{eqnarray*}\hfill} \hfill\parbox[t][1cm][b]{1cm}
{\hfill \begin{eqnarray}\end{eqnarray}}
\noindent\parbox[t][1cm][b]{8cm}{
\begin{eqnarray*}
{[}\gamma_\rho\gamma_\mu\gamma_\sigma\gamma_{L,R}{]}_1
{[}\gamma_\sigma\gamma_\nu\gamma_\rho\gamma_{L,R}{]}_2
                   &=&4{[}\gamma_\nu\gamma_{L,R}{]}_1
                       {[}\gamma_\mu\gamma_{L,R}{]}_2
                 +(n-4){[}\gamma_\mu\gamma_{L,R}{]}_1
                       {[}\gamma_\nu\gamma_{L,R}{]}_2
\end{eqnarray*}\hfill} \hfill\parbox[t][1cm][b]{1cm}
{\hfill \begin{eqnarray}\end{eqnarray}}
\noindent\parbox[t][1cm][b]{8cm}{
\begin{eqnarray*}
{[}\gamma_\rho\gamma_\mu\gamma_\sigma\gamma_{L,R}{]}_1
{[}\gamma_\sigma\gamma_\nu\gamma_\rho\gamma_{R,L}{]}_2
&=&4\delta_{\mu\nu}{[}\gamma_\alpha\gamma_{L,R}{]}_1
                       {[}\gamma_\alpha\gamma_{R,L}{]}_2
                 +(n-4){[}\gamma_\mu\gamma_{L,R}{]}_1
                       {[}\gamma_\nu\gamma_{R,L}{]}_2
\end{eqnarray*}\hfill} \hfill\parbox[t][1cm][c]{1cm}
{\hfill \begin{eqnarray}\label{eq:fourthBoxid}\end{eqnarray}}

\noindent where the square brackets
$[\ ]_1$ and $[\ ]_2$ indicate that the enclosed
$\gamma$-matrices are associated with the external fermion currents
$J_1$ and $J_2$ respectively.

The general ${\cal O}(N_f\alpha^2)$ box diagram for massless external
fermions at $q=0$ therefore takes the form
\begin{equation}
B_{\mu\nu}J_{1\mu}J_{2\nu}=B\cdot J_{1\alpha}J_{2\alpha}.
\end{equation}
Here $B_{\mu\nu}$ is a tensor integral and the product of
$J_{1\alpha}J_{2\alpha}$ can be constructed from one or a combination of
the $\gamma$-matrices appearing in
eq.(\ref{eq:firstBoxid})--(\ref{eq:fourthBoxid}). As for the self-energy
contributions, the tensor integral $B_{\mu\nu}$, can only be proportional
to $\delta_{\mu\nu}$ and hence, using the projection operator method of
section~\ref{sect:WSelfe},
the scalar integral representation for the form factor $B$ is
seen to be $B=(\delta_{\mu\nu}/n)B_{\mu\nu}=B_{\mu\mu}/n$.
In this case the identities
(\ref{eq:firstBoxid})--(\ref{eq:fourthBoxid}) simplify to become
\begin{eqnarray}
\frac{\delta_{\mu\nu}}{n}
{[}\gamma_\rho\gamma_\mu\gamma_\sigma\gamma_{L,R}{]}_1
{[}\gamma_\rho\gamma_\nu\gamma_\sigma\gamma_{L,R}{]}_2
&=&\frac{(5n-4)}{n}{[}\gamma_\alpha\gamma_{L,R}{]}_1
                   {[}\gamma_\alpha\gamma_{L,R}{]}_2\\
\frac{\delta_{\mu\nu}}{n}
{[}\gamma_\rho\gamma_\mu\gamma_\sigma\gamma_{L,R}{]}_1
{[}\gamma_\rho\gamma_\nu\gamma_\sigma\gamma_{R,L}{]}_2
                   &=&{[}\gamma_\alpha\gamma_{L,R}{]}_1
                       {[}\gamma_\alpha\gamma_{R,L}{]}_2
\label{eq:secondBoxid}\\
\frac{\delta_{\mu\nu}}{n}
{[}\gamma_\rho\gamma_\mu\gamma_\sigma\gamma_{L,R}{]}_1
{[}\gamma_\sigma\gamma_\nu\gamma_\rho\gamma_{L,R}{]}_2
                   &=&{[}\gamma_\alpha\gamma_{L,R}{]}_1
                       {[}\gamma_\alpha\gamma_{L,R}{]}_2\\
\frac{\delta_{\mu\nu}}{n}
{[}\gamma_\rho\gamma_\mu\gamma_\sigma\gamma_{L,R}{]}_1
{[}\gamma_\sigma\gamma_\nu\gamma_\rho\gamma_{R,L}{]}_2
&=&\frac{(5n-4)}{n}{[}\gamma_\alpha\gamma_{L,R}{]}_1
                   {[}\gamma_\alpha\gamma_{R,L}{]}_2.
\end{eqnarray}

When the diagrams of Fig.\ref{fig:BoxDiagrams} are combined with 1-loop
box diagrams, in which 1-loop fermionic $\overline{{\rm MS}}$ counterterms
have been inserted in all possible ways, the result was found to be
finite and give a contribution to $\Delta r^{(2)}$ of
\begin{multline}
\Delta r^{(2)}_{{\rm box}}=-2\left(\frac{g^2}{16\pi^2}\right)^2\bigg\{
      \frac{(3-8s_\theta^2)(1-2s_\theta^2)}{3c_\theta^2}
            \left(\ln\frac{M_Z^2}{\mu^{\prime2}}-\frac{5}{3}\right)
     -\frac{(21-52 s_\theta^2)}{9s_\theta^2}\ln c_\theta^2\\
     +\frac{(3-7s_\theta^2)}{3s_\theta^2}
            \left(\ln c_\theta^2+2\ln\frac{M_Z^2}{\mu^{\prime2}}\right)
                 \ln c_\theta^2
       \bigg\}
\label{eq:DeltarBox}
\end{multline}

\section{Conclusions}

Combining the contributions from $W$ self energy corrections given in
Eq.(\ref{eq:DeltarSEH}) and Eq.(\ref{eq:DeltarSEW}) with
the vertex corrections of Eq.(\ref{eq:DeltarVertex}) and box diagram
corrections of Eq.(\ref{eq:DeltarBox}) for one complete massless
generation of fermions finally gives
\begin{eqnarray}
\Delta r^{(2)}&=&
-\left(\frac{g^2}{16\pi^2}\right)^2\bigg\{\frac{5}{2s_h^2}
    -\frac{2}{s_h^2}\ln\frac{M_W^2}{\mu^{\prime2}}
    -\frac{\ln c_h^2}{s_h^4}
          \left(\ln\frac{M_W^2}{\mu^{\prime2}}
               +\ln\frac{M_H^2}{\mu^{\prime2}}-\frac{5}{2}\right)
\nonumber\\ & &\qquad\qquad
-\frac{(369-878s_\theta^2+334s_\theta^4+160s_\theta^6)}{72c_\theta^4}
+\frac{(7-14s_\theta^2+12s_\theta^4)}{12c_\theta^4}\pi^2
\nonumber\\ & &\qquad\qquad
-\frac{(57-40s_\theta^2)}{9s_\theta^2}\ln c_\theta^2
+\frac{(3+2s_\theta^2)}{3s_\theta^2}
      \left(\ln c_\theta^2+2\ln \frac{M_Z^2}{\mu^{\prime2}}\right)
      \ln c_\theta^2
\\ & &\qquad\qquad
-\frac{(5-6s_\theta^2+22s_\theta^4-16s_\theta^6)}{6c_\theta^4}
      \ln\frac{M_Z^2}{\mu^{\prime2}}
+\frac{(6-12s_\theta^2+11s_\theta^4)}{3c_\theta^4}
      \ln^2\frac{M_Z^2}{\mu^{\prime2}}
\bigg\}\nonumber
\end{eqnarray}
which obviously simplifies substantially for $\mu^\prime=M_Z$.
Upon evaluation one finds, for a 't~Hooft mass $\mu^\prime=91.1867$\,GeV
that $\Delta r^{(2)}=-5.45\times10^{-5}$,
$-7.28\times10^{-5}$ and $-1.54\times10^{-4}$ for $M_H=$100\,GeV,
300\,GeV and 1000\,GeV respectively. $s_\theta$ was set to its
corresponding $\overline{{\rm MS}}$ values of 0.2316, 0.2322 and 0.2330
obtained from the program {\tt Z0POLE} \cite{Z0POLE}.

For physics on or above the $Z^0$ resonance, the number of light
generations is at least 2 and could be taken to be 3 depending on
exactly how the top quark mass corrections are to be treated.
The correction is large compared with what would be expected for
a 2-loop electroweak correction and therefore clearly displays the
enhancement with the fermion number, $N_f$.

As mentioned in the introduction, a number of classes
of 2-loop contributions to $\Delta r$ have now been computed. Few,
if any, dominant classes remain to be tackled and the next logical
step is the complete set of 2-loop corrections. In order to use these
in making theoretical predictions the full renormalization of the
Standard Model at 2-loop order
would be required. Many of the issues that need to
be confronted in undertaking the full renormalization have been
encountered in this work and in ref.\cite{MaldeStuart1}.

\section*{Acknowledgments}

RGS wishes to thank the Max-Planck-Institut f\"ur Physik, Munich,
for hospitality while part of this work was carried out.
This work was supported in part by the US Department of Energy.

\appendix

\section{The Master Integral}
\label{sect:MasterInt}

If the general coupling of a fermion to a vector boson will
be denoted
\[
\begin{picture}(43,42)(0,0)
\ArrowLine(0,-36)(12,0)
\ArrowLine(12,0)(0,36)
\Photon(12,0)(40,0){4}{3}
\Vertex(12,0){1}
\end{picture}
\equiv i\gamma_\mu(\beta_L\gamma_L+\beta_R\gamma_R)
\]\newline
\newline
\noindent where $\gamma_L$ and $\gamma_R$ are the usual left- and right-handed
helicity projection operators and $\beta_L$ and $\beta_R$ are the
corresponding coupling constants then the
${\cal O}(N_f\alpha^2)$ diagrams that are obtained
from 1-loop diagrams by the insertion of a massless fermion loop
into an internal vector boson line are most easily calculated
by means of the following identity.
For the massless fermion loop insertion, it may be shown that
\begin{eqnarray}
\begin{picture}(32,5)(26,0)
\Photon(0,0)(22,0){4}{2}
\ArrowArc(31,0)(9,0,180)
\ArrowArc(31,0)(9,180,0)
\Vertex(22,0){1}
\Vertex(40,0){1}
\Photon(40,0)(62,0){4}{2}
\end{picture}
&=&-\left(\delta_{\mu\nu}-\frac{p_\mu p_\nu}{p^2}\right)
   \frac{(\beta_L\beta_L^\prime+\beta_R\beta_R^\prime)}{16\pi^2}
   \frac{\displaystyle (n-2)}{\displaystyle (n-1)}
   \int\frac{d^n q}{\displaystyle i\pi^2}\frac{p^2}{q^2(q+p)^2}\\
&=&-\left(p^2\delta_{\mu\nu}-p_\mu p_\nu\right)
   \frac{(\beta_L\beta_L^\prime+\beta_R\beta_R^\prime)}{16\pi^2}
   4(\pi p^2)^{\frac{n}{2}-2}
             \frac{\Gamma^2\!\left(\frac{\displaystyle n}
                                        {\displaystyle 2}\right)}
                  {\Gamma(n)}
             \Gamma\left(2-\frac{\displaystyle n}
                                {\displaystyle 2}\right)
\end{eqnarray}
where $\beta_L$, $\beta_R$ and $\beta_L^\prime$, $\beta_R^\prime$ are the
couplings of the attached vector bosons.

The master integral, to which all ${\cal O}(N_f\alpha^2)$ diagrams
relevant for the present calculation can be reduced, takes the form
\begin{equation}
I(j,k,l,m,n,M^2)=
\int\frac{d^np}{i\pi^2}\frac{1}{[p^2]^j[p^2+M^2]^k}
\int\frac{d^nq}{i\pi^2}\frac{1}{[q^2]^l[(q+p)^2]^m}.
\label{eq:MasterIntDef}
\end{equation}

In eq.(\ref{eq:MasterIntDef}), the integration over $q$ can be performed
using standard Feynman parameter techniques and yields
\begin{equation}
\int\frac{d^nq}{i\pi^2}\frac{1}{[q^2]^l[(q+p)^2]^m}=
\frac{\pi^{\frac{n}{2}-2}}{[p^2]^{l+m-\frac{n}{2}}}
\frac{\Gamma\left(l+m-\frac{\textstyle n}{\textstyle 2}\right)
      \Gamma\left(\frac{\textstyle n}{\textstyle 2}-l\right)
      \Gamma\left(\frac{\textstyle n}{\textstyle 2}-m\right)}
     {\Gamma(l)\Gamma(m)\Gamma(n-l-m)}.
\end{equation}
The resulting integral with respect to $p$ in eq.(\ref{eq:MasterIntDef})
is independent of angle and hence
\begin{align}
\int\frac{d^np}{i\pi^2}\ \frac{1}{[p^2]^{j+l+m-\frac{n}{2}}[p^2+M^2]^k}
=\frac{2\pi^{\frac{n}{2}-2}}
        {\Gamma\left(\frac{\textstyle n}{\textstyle 2}\right)}
  &\int_0^\infty dp\ \frac{p^{2n-2j-2l-2m-1}}{[p^2+M^2]^k}\nonumber\\
=\frac{\pi^{\frac{n}{2}-2}}{(M^2)^{k+j+l+m-n}}
  &\frac{\Gamma(n-j-l-m)\Gamma(k+j+l+m-n)}
        {\Gamma\left(\frac{\textstyle n}{\textstyle 2}\right)\Gamma(k)}
\nonumber\\
\end{align}
from which it follows
\begin{eqnarray}
 & &I_0(j,k,l,m,n,M^2)=
    \frac{\pi^{n-4}}{(M^2)^{k+j+l+m-n}}\nonumber\\
 & &\ \ \ \times
    \frac{\Gamma(n-j-l-m)\Gamma(k+j+l+m-n)
          \Gamma\left(l+m-\frac{\textstyle n}{\textstyle 2}\right)
          \Gamma\left(\frac{\textstyle n}{\textstyle 2}-l\right)
          \Gamma\left(\frac{\textstyle n}{\textstyle 2}-m\right)}
         {\Gamma\left(\frac{\textstyle n}{\textstyle 2}\right)
          \Gamma(k)\Gamma(l)\Gamma(m)\Gamma(n-l-m)}
\nonumber\\
\label{eq:MasterIntexpr}
\end{eqnarray}

For the purposes of compactness it is useful to define the related
integrals
\begin{eqnarray}
I_1(j,n,M_1^2,M_2^2)&=&
\int\frac{d^np}{i\pi^2}\frac{1}{[p^2]^j[p^2+M_1^2][p^2+M_2^2]}
\int\frac{d^nq}{i\pi^2}\frac{1}{q^2(q+p)^2}\\
 &=&\frac{1}{(M_1^2-M_2^2)}\left\{I_0(j,1,1,1,n,M_2^2)
                               -I_0(j,1,1,1,n,M_1^2)\right\}
\label{eq:MasterInt1Def}
\end{eqnarray}
and
\begin{eqnarray}
I_2(j,n,M_1^2,M_2^2)&=&
\int\frac{d^np}{i\pi^2}\frac{1}{[p^2]^j[p^2+M_1^2][p^2+M_2^2]^2}
\int\frac{d^nq}{i\pi^2}\frac{1}{q^2(q+p)^2}\\
 &=&\frac{1}{(M_1^2-M_2^2)^2}\{I_0(j,1,1,1,n,M_1^2)
\nonumber\\ & &\qquad   +(M_1^2-M_2^2)I_0(j,2,1,1,n,M_2^2)
                                   -I_0(j,1,1,1,n,M_2^2)\}\nonumber\\
\label{eq:MasterInt2Def}
\end{eqnarray}
Finally, the $q\neq0$ amplitudes were obtained independently and required
varied and extensive techniques and will appear in a separate publication
\cite{MaldeThesis}.

\section{$W$ Self Energy Corrections}

In this appendix the contributions from individual Feynman diagrams to
the ${\cal O}(N_f\alpha^2)$ $W$ boson self-energy, $\Pi_{WW}^{(2)}(0)$,
are listed. Their net effect on the inverse muon lifetime,
$\Gamma^{(0)}=g^4m_\mu^5/(6144\pi^3M_W^4)$, is to induce a shift of
$\Delta\Gamma^{(2)}=2\Gamma^{(0)}\Pi_{WW}^{(2)}(0)/M_W^2$
or equivalently produce a contribution of $\Pi_{WW}^{(2)}(0)/M_W^2$
to $\Delta r^{(2)}$.
Also listed are the divergent parts of the diagrams at general $q^2$.
The diagrams are labeled according to Fig.\ref{fig:WSelfe}.
Thus $\Pi_{WW}^{(2{\rm a})}(0)$ denotes the contribution from diagram
of Fig.\ref{fig:WSelfe}(a) at $q^2=0$ and
$\Delta\Pi_{WW}^{(2{\rm a})}(q^2)$ denotes its divergent part at general
$q^2$ with $\epsilon=2-n/2$.
In the following expressions, an overall common factor of $\left(g^2/(16\pi)^2\right)^2\delta_{\mu\nu}$
has been omitted for brevity and
\begin{equation}
{\tilde B_0}({q^2},{M^{2}_{1}},{M^{2}_{2}})
\equiv -\int_0^1 \ln\left(-q^2 x^2+(q^2-M_1^2+M_2^2) x
                          +M_1^2-i\epsilon\right)\,dx.
\nonumber
\end{equation}

\begin{align}
\intertext{{\bf Diagram (a)}}
\intertext{Internal photon}
\Pi_{WW}^{(2\,{\rm a})}(0)=&-10s_\theta^2
                            \frac{(n-2)}{n}I_0(-1,2,1,1,n,M_W^2)\\
\Delta\Pi_{WW}^{(2\,{\rm a})}(q^2)=& {\frac{{s^{2}_{\theta}}}{9\,{{\epsilon}^{2}}}}\{
  45\,{M^{2}_{W}} - 22\,{q^2}\}
\nonumber \\
&  \hspace{\eqnoffstc} +{\frac{{s^{2}_{\theta}}}{54\,\epsilon\,{q^4}}}\Big\{
  240\,{M^{4}_{W}}\,{q^2}
 +1077\,{M^{2}_{W}}\,{q^4}
 -442\,{q^6}
\nonumber \\
&  \hspace{\eqnoffst} + [ 240\,{M^{6}_{W}} + 792\,{M^{4}_{W}}\,{q^2} -
    540\,{M^{2}_{W}}\,{q^4} ] \,
  {\ln{M^{2}_{W}}}
\nonumber \\
&  \hspace{\eqnoffst} -[ 240\,{M^{6}_{W}} +
       792\,{M^{4}_{W}}\,{q^2} - 264\,{q^6} ] \,
     {\ln({q^2}+{M^{2}_{W}})} \Big\}
\nonumber \\
\intertext{Internal $Z^0$}
\Pi_{WW}^{(2\,{\rm a})}(0)=&-10c_\theta^2
                            \frac{(n-2)}{n}I_2(-2,n,M_Z^2,M_W^2)\\
\Delta\Pi_{WW}^{(2\,{\rm a})}(q^2)=& {-\frac{1}{18\,{{\epsilon}^{2}}}}\{
  44\,{c^{2}_{\theta}}\,{q^2} -
   45\,{M^{2}_{W}}\,
    ( 3 - 2\,{s^{2}_{\theta}} ) \}
\nonumber \\
&  \hspace{\eqnoffstc} +{\frac{1}{\epsilon\,( 216\,{c^{4}_{\theta}}\,
         ( 1 + {c^{2}_{\theta}} ) \,
         {M^{2}_{W}}\,{q^4} +
        108\,{c^{6}_{\theta}}\,{q^6} +
        108\,{c^{2}_{\theta}}\,{M^{4}_{W}}\,{q^2}\,
         {s^{4}_{\theta}} ) }}
\nonumber \\
& \hspace{\eqnoffsta} \times \Big\{
  172\,{c^{8}_{\theta}}\,{q^8}
+{c^{6}_{\theta}}\,{M^{2}_{W}}\,{q^6}\,
  ( 2011 + 48\,{\ln{c^2_{\theta} }} -
    1442\,{s^{2}_{\theta}} )
\nonumber \\
&  \hspace{\eqnoffst} +12\,{M^{8}_{W}}\,{s^{4}_{\theta}}\,
  ( 33 - 40\,{s^{2}_{\theta}} ) \,
  ( {\ln{c^2_{\theta} }} + {s^{2}_{\theta}}
     )
\nonumber \\
&  \hspace{\eqnoffst} +2\,{c^{4}_{\theta}}\,{M^{4}_{W}}\,{q^4}\,
  ( 1350 + 6\,{\ln{c^2_{\theta} }} -
    1449\,{s^{2}_{\theta}} + 368\,{s^{4}_{\theta}}
     )
\nonumber \\
&  \hspace{\eqnoffst} +3\,{c^{2}_{\theta}}\,{M^{6}_{W}}\,{q^2}\,
  [ 8\,{\ln{c^2_{\theta} }}\,
     ( 78 - 149\,{s^{2}_{\theta}} +
       66\,{s^{4}_{\theta}} )  +
    {s^{2}_{\theta}}\,
     ( 624 - 943\,{s^{2}_{\theta}} +
       338\,{s^{4}_{\theta}} )  ]
\nonumber \\
&  \hspace{\eqnoffst} -[ 12\,{M^{8}_{W}}\,{s^{6}_{\theta}}\,
     ( 33 - 40\,{s^{2}_{\theta}} )  +
    24\,{c^{6}_{\theta}}\,{M^{2}_{W}}\,{q^6}\,
     ( 25 - 23\,{s^{2}_{\theta}} )
\nonumber \\
&  \hspace{\eqnoffstb} +
    12\,{c^{4}_{\theta}}\,{M^{4}_{W}}\,{q^4}\,
     ( 92 - 95\,{s^{2}_{\theta}} +
       4\,{s^{4}_{\theta}} )
\nonumber \\
&  \hspace{\eqnoffstb} + 24\,{M^{6}_{W}}\,{q^2}\,{s^{2}_{\theta}}\,
     ( 72 - 190\,{s^{2}_{\theta}} +
       159\,{s^{4}_{\theta}} - 41\,{s^{6}_{\theta}}
        )  ] \,{\ln{M^{2}_{W}}}
\nonumber \\
&  \hspace{\eqnoffst} -[ 528\,{c^{8}_{\theta}}\,{q^8} +
     12\,{c^{6}_{\theta}}\,{M^{2}_{W}}\,{q^6}\,
      ( 91 - 44\,{s^{2}_{\theta}} )  +
     12\,{M^{8}_{W}}\,{s^{6}_{\theta}}\,
      ( 33 - 40\,{s^{2}_{\theta}} )
\nonumber \\
&  \hspace{\eqnoffstb} - 12\,{c^{4}_{\theta}}\,{M^{4}_{W}}\,{q^4}\,
      ( 448 - 535\,{s^{2}_{\theta}} +
        132\,{s^{4}_{\theta}} )
\nonumber \\
&  \hspace{\eqnoffstb} + 12\,{M^{6}_{W}}\,{q^2}\,{s^{2}_{\theta}}\,
      ( 144 - 515\,{s^{2}_{\theta}} +
        543\,{s^{4}_{\theta}} - 172\,{s^{6}_{\theta}}
         )  ] \,{{\tilde B_0}({q^2},{M^{2}_{W}},
    {M^{2}_{Z}})}\Big\}
\nonumber \\
\intertext{\bf Diagram (b)}
\Pi_{WW}^{(2\,{\rm b})}(0)=&2\frac{(n-2)(n-1)}{n}I_0(-1,2,1,1,n,M_W^2)\\
\Delta\Pi_{WW}^{(2\,{\rm b})}(q^2)=&  {-\frac{3\,{M^{2}_{W}}}{{{\epsilon}^{2}}}}
 -{\frac{{M^{2}_{W}}}{2\,\epsilon}}\{5
-12\,{\ln{M^{2}_{W}}}\}
\nonumber \\
\intertext{{\bf Diagram (c)}}
\Pi_{WW}^{(2\,{\rm c})}(0)=&-2M_W^2\frac{(n-2)}{n}I_2(-1,n,M_H^2,M_W^2)\\
\Delta\Pi_{WW}^{(2\,{\rm c})}(q^2)=& {-\frac{{M^{2}_{W}}}{2\,{{\epsilon}^{2}}}}
\nonumber \\
&  \hspace{\eqnoffstc} -{\frac{{M^{2}_{W}}}
    {\epsilon\,( 24\,{c^{4}_{h}}\,
         ( 1 + {c^{2}_{h}} ) \,
         {M^{2}_{W}}\,{q^4} +
        12\,{c^{6}_{h}}\,{q^6} +
        12\,{c^{2}_{h}}\,{M^{4}_{W}}\,{q^2}\,
         {{{s_{h}}}^4} ) }}
\nonumber \\
&  \hspace{\eqnoffsta} \times \Big\{5\,{c^{6}_{h}}\,{q^6}
 -2\,{c^{4}_{h}}\,{M^{2}_{W}}\,{q^4}\,
  ( 6 + 2\,{\ln{c^2_{h} }} - 9\,{{{s_{h}}}^2} )
-4\,{M^{6}_{W}}\,{{{s_{h}}}^4}\,
  ( {\ln{c^2_{h} }} + {{{s_{h}}}^2} )
\nonumber \\
&  \hspace{\eqnoffst} + {c^{2}_{h}}\,{M^{4}_{W}}\,{q^2}\,
    ( 16\,{\ln{c^2_{h} }} + 16\,{{{s_{h}}}^2} -
      24\,{{{s_{h}}}^2}\,{\ln{c^2_{h} }} - 19\,{{{s_{h}}}^4}
       )
\nonumber \\
&  \hspace{\eqnoffst} + [ 8\,{c^{2}_{h}}\,{M^{4}_{W}}\,{q^2}\,
       {{{s_{h}}}^4} + 4\,{M^{6}_{W}}\,{{{s_{h}}}^6} +
      4\,{c^{4}_{h}}\,{M^{2}_{W}}\,{q^4}\,
       ( 4 - 3\,{{{s_{h}}}^2} )  ] \,
    {\ln{M^{2}_{W}}}
\nonumber \\
&  \hspace{\eqnoffst} +[ 12\,{c^{6}_{h}}\,{q^6} +
     20\,{c^{2}_{h}}\,{M^{4}_{W}}\,{q^2}\,
      {{{s_{h}}}^4} + 4\,{M^{6}_{W}}\,{{{s_{h}}}^6}
\nonumber \\
&  \hspace{\eqnoffstb} + 4\,{c^{4}_{h}}\,{M^{2}_{W}}\,{q^4}\,
      ( 16 - 9\,{{{s_{h}}}^2} )  ] \,
   {{\tilde B_0}({q^2},{M^{2}_{H}},{M^{2}_{W}}
    )}\Big\}
\nonumber \\
\intertext{\bf Diagram (d)}
\intertext{Internal photon}
\Pi_{WW}^{(2\,{\rm d})}(0)=&-\frac{80}{3}s_\theta^4
                            \frac{(n-2)}{n}I_0(0,1,1,1,n,M_W^2)\\
\Delta\Pi_{WW}^{(2\,{\rm d})}(q^2)=&  {\frac{{s^{4}_{\theta}}}{27\,{{\epsilon}^{2}}}}\{
  180\,{M^{2}_{W}} - 176\,{q^2}\}
\nonumber \\
&  \hspace{\eqnoffstc} +{\frac{{s^{4}_{\theta}}}{81\,\epsilon\,{q^4}}}\Big\{
  168\,{M^{4}_{W}}\,{q^2}
+2418\,{M^{2}_{W}}\,{q^4}
 -1768\,{q^6}
\nonumber \\
&  \hspace{\eqnoffst} +[ 168\,{M^{6}_{W}} + 1152\,{M^{4}_{W}}\,{q^2} -
    1152\,{M^{2}_{W}}\,{q^4} ] \,
  {\ln{M^{2}_{W}}}
\nonumber \\
&  \hspace{\eqnoffst} - [ 168\,{M^{6}_{W}} +
       1152\,{M^{4}_{W}}\,{q^2} -
       72\,{M^{2}_{W}}\,{q^4} - 1056\,{q^6} ] \,
     {\ln({q^2} + {M^{2}_{W}})} \Big\}
\nonumber\\
\intertext{Internal $Z^0$}
\Pi_{WW}^{(2\,{\rm d})}(0)=&-10
                      \left(1-2s_\theta^2+\frac{8}{3}s_\theta^4\right)
                      \frac{(n-2)}{n}I_2(-2,n,M_W^2,M_Z^2)\\
\Delta\Pi_{WW}^{(2\,{\rm d})}(q^2)=&  {-\frac{(3 - 6\,{s^{2}_{\theta}} + 8\,{s^{4}_{\theta}})}
    {54\,{c^{2}_{\theta}}\,{{\epsilon}^{2}}}}\{
  44\,{c^{2}_{\theta}}\,{q^2} -
   45\,{M^{2}_{W}}\,
    ( 3 - {s^{2}_{\theta}} ) \}
\nonumber \\
&  \hspace{\eqnoffstc} +{\frac{(3 - 6\,{s^{2}_{\theta}} + 8\,{s^{4}_{\theta}})}
    {\epsilon\,( 648\,{c^{6}_{\theta}}\,
         ( 1 + {c^{2}_{\theta}} ) \,
         {M^{2}_{W}}\,{q^4} +
        324\,{c^{8}_{\theta}}\,{q^6} +
        324\,{c^{4}_{\theta}}\,{M^{4}_{W}}\,{q^2}\,
         {s^{4}_{\theta}} ) }}
\nonumber \\
&  \hspace{\eqnoffsta}\times \Big\{
  172\,{c^{8}_{\theta}}\,{q^8}
+{c^{6}_{\theta}}\,{M^{2}_{W}}\,{q^6}\,
  ( 2011 + 552\,{\ln{c^2_{\theta} }} -
    569\,{s^{2}_{\theta}} )
\nonumber \\
&  \hspace{\eqnoffst} -12\,{M^{8}_{W}}\,{s^{4}_{\theta}}\,
  ( {\ln{c^2_{\theta} }} + {s^{2}_{\theta}}
     ) \,( 33 + 7\,{s^{2}_{\theta}} )
\nonumber \\
&  \hspace{\eqnoffst} +2\,{c^{4}_{\theta}}\,{M^{4}_{W}}\,{q^4}\,
  ( 1350 - 1251\,{s^{2}_{\theta}} +
    269\,{s^{4}_{\theta}} +
    6\,{\ln{c^2_{\theta} }}\,
     ( 91 - 87\,{s^{2}_{\theta}} )  )
\nonumber \\
&  \hspace{\eqnoffst} -3\,{c^{2}_{\theta}}\,{M^{6}_{W}}\,{q^2}\,
  ( 8\,{\ln{c^2_{\theta} }}\,
     ( 78 - 13\,{s^{2}_{\theta}} -
       24\,{s^{4}_{\theta}} )  +
    {s^{2}_{\theta}}\,
     ( 624 - 305\,{s^{2}_{\theta}} +
       19\,{s^{4}_{\theta}} )  )
\nonumber \\
&  \hspace{\eqnoffst} -[ 24\,{c^{6}_{\theta}}\,{M^{2}_{W}}\,{q^6}\,
     ( 25 - 2\,{s^{2}_{\theta}} )  -
    12\,{M^{8}_{W}}\,{s^{6}_{\theta}}\,
     ( 33 + 7\,{s^{2}_{\theta}} )
\nonumber \\
&  \hspace{\eqnoffstb} +    12\,{c^{4}_{\theta}}\,{M^{4}_{W}}\,{q^4}\,
     ( 92 - 89\,{s^{2}_{\theta}} +
       {s^{4}_{\theta}} )
\nonumber \\
&  \hspace{\eqnoffstb} -
    24\,{M^{6}_{W}}\,{q^2}\,{s^{2}_{\theta}}\,
     ( 72 - 98\,{s^{2}_{\theta}} +
       21\,{s^{4}_{\theta}} + 5\,{s^{6}_{\theta}}
        )  ] \,{\ln{M^{2}_{W}}}
\nonumber \\
&  \hspace{\eqnoffst} -[ 528\,{c^{8}_{\theta}}\,{q^8} +
     12\,{c^{6}_{\theta}}\,{M^{2}_{W}}\,{q^6}\,
      ( 91 - 47\,{s^{2}_{\theta}} )  -
     12\,{M^{8}_{W}}\,{s^{6}_{\theta}}\,
      ( 33 + 7\,{s^{2}_{\theta}} )
\nonumber \\
&  \hspace{\eqnoffstb} -
     12\,{c^{4}_{\theta}}\,{M^{4}_{W}}\,{q^4}\,
      ( 448 - 361\,{s^{2}_{\theta}} +
        45\,{s^{4}_{\theta}} )
\nonumber \\
&  \hspace{\eqnoffstb} -  12\,{M^{6}_{W}}\,{q^2}\,{s^{2}_{\theta}}\,
      ( 144 - 61\,{s^{2}_{\theta}} -
        138\,{s^{4}_{\theta}} + 55\,{s^{6}_{\theta}}
         )  ] \,{{\tilde B_0}({q^2},{M^{2}_{W}},
    {M^{2}_{Z}})}\Big\}
\nonumber\\
\intertext{Internal $Z$-$\gamma$ mixing}
\Pi_{WW}^{(2\,{\rm d})}(0)=&-20 s_\theta^2
                            \left(1-\frac{8}{3}s_\theta^2\right)
                            \frac{(n-2)}{n}I_1(-1,n,M_W^2,M_Z^2)\\
\Delta\Pi_{WW}^{(2\,{\rm d})}(q^2)=&  {-\frac{{s^{2}_{\theta}}\,
      ( 3 - 8\,{s^{2}_{\theta}} ) }{27\,
      {c^{2}_{\theta}}\,{{\epsilon}^{2}}}}\{
  44\,{c^{2}_{\theta}}\,{q^2} -
   45\,{M^{2}_{W}}\,
    ( 2 - {s^{2}_{\theta}} ) \}
\nonumber \\
&  \hspace{\eqnoffstc} +{\frac{{s^{2}_{\theta}}\,
      ( 3 - 8\,{s^{2}_{\theta}} ) }{162\,
      {c^{4}_{\theta}}\,\epsilon\,{M^{2}_{W}}\,{q^4}}}
  \Big\{240\,{c^{6}_{\theta}}\,{q^8}
+4\,{c^{4}_{\theta}}\,{M^{2}_{W}}\,{q^6}\,
  ( 67 - 30\,{\ln{c^2_{\theta} }} -
    114\,{s^{2}_{\theta}} )
\nonumber \\
&  \hspace{\eqnoffst} -12\,{M^{6}_{W}}\,{q^2}\,
  ( 1 - ( 3 - 11\,{\ln{c^2_{\theta} }} ) \,
     {s^{2}_{\theta}} +
    ( 14 - {\ln{c^2_{\theta} }} ) \,
     {s^{4}_{\theta}} - 2\,{s^{6}_{\theta}} )
\nonumber \\
&  \hspace{\eqnoffst} +3\,{c^{2}_{\theta}}\,{M^{4}_{W}}\,{q^4}\,
  ( 300 - 255\,{s^{2}_{\theta}} +
    64\,{s^{4}_{\theta}} +
    4\,{\ln{c^2_{\theta} }}\,
     ( 2 + 9\,{s^{2}_{\theta}} )  )
\nonumber \\
&  \hspace{\eqnoffst} -[ 12\,{c^{6}_{\theta}}\,{M^{8}_{W}} -
    120\,{c^{4}_{\theta}}\,{M^{2}_{W}}\,{q^6}\,
     ( 2 - {s^{2}_{\theta}} )
\nonumber \\
&  \hspace{\eqnoffstb}  - 12\,{M^{6}_{W}}\,{q^2}\,
     ( 7 - 21\,{s^{2}_{\theta}} +
       32\,{s^{4}_{\theta}} - 8\,{s^{6}_{\theta}}
        )
\nonumber \\
&  \hspace{\eqnoffstb} + 12\,{M^{4}_{W}}\,{q^4}\,
     ( 7 + 16\,{s^{2}_{\theta}} -
       42\,{s^{4}_{\theta}} + 19\,{s^{6}_{\theta}}
        )  ] \,{\ln{M^{2}_{W}}}
\nonumber \\
&  \hspace{\eqnoffst} + [ 12\,{c^{6}_{\theta}}\,{M^{8}_{W}} -
      84\,{c^{6}_{\theta}}\,{M^{6}_{W}}\,{q^2} -
      324\,{c^{6}_{\theta}}\,{M^{4}_{W}}\,{q^4} -
      348\,{c^{6}_{\theta}}\,{M^{2}_{W}}\,{q^6}
\nonumber \\
&  \hspace{\eqnoffstb} -
      120\,{c^{6}_{\theta}}\,{q^8} ] \,
    {\ln({q^2} + {M^{2}_{W}})}
\nonumber \\
&  \hspace{\eqnoffst} -[ 120\,{c^{6}_{\theta}}\,{q^8} +
       12\,{c^{4}_{\theta}}\,{M^{2}_{W}}\,{q^6}\,
        ( 53 - 19\,{s^{2}_{\theta}} )  -
       12\,{M^{6}_{W}}\,{q^2}\,{s^{4}_{\theta}}\,
        ( 11 - {s^{2}_{\theta}} )
\nonumber \\
&  \hspace{\eqnoffstb}  -
       24\,{M^{4}_{W}}\,{q^4}\,
        ( 28 - 35\,{s^{2}_{\theta}} +
          3\,{s^{4}_{\theta}} + 4\,{s^{6}_{\theta}}
           )  ] \,{{\tilde B_0}({q^2},{M^{2}_{W}},
      {M^{2}_{Z}})} \Big\}
\nonumber\\
\intertext{\bf Diagram (e)}
\intertext{Internal photon}
\Pi_{WW}^{(2\,{\rm e})}(0)=&0\\
\Delta\Pi_{WW}^{(2\,{\rm e})}(q^2)=&0\\
\intertext{Internal $Z^0$}
\Pi_{WW}^{(2\,{\rm e})}(0)=&2\left(1-2s_\theta^2+\frac{8}{3}s_\theta^4\right)
                            \frac{(n-1)(n-2)}{n}I_0(-1,2,1,1,n,M_Z^2)\\
\Delta\Pi_{WW}^{(2\,{\rm e})}(q^2)=& {-\frac{{M^{2}_{W}}\,
      ( 3 - 6\,{s^{2}_{\theta}} +
        8\,{s^{4}_{\theta}} ) }{{c^{2
       }_{\theta}}\,{{\epsilon}^{2}}}}
\nonumber \\
&  \hspace{\eqnoffst} -{\frac{{M^{2}_{W}}\,
      ( 3 - 6\,{s^{2}_{\theta}} +
        8\,{s^{4}_{\theta}} ) }{6\,
      {c^{2}_{\theta}}\,\epsilon}}\{5
+12\,{\ln{c^2_{\theta} }}
-12\,{\ln{M^{2}_{W}}}\}
\nonumber\\
\intertext{Internal $Z$-$\gamma$ mixing}
\Pi_{WW}^{(2\,{\rm e})}(0)=&4 s_\theta^2
                            \left(1-\frac{8}{3}s_\theta^2\right)
                            \frac{(n-1)(n-2)}{n}I_0(0,1,1,1,n,M_Z^2)\\
\Delta\Pi_{WW}^{(2\,{\rm e})}(q^2)=& {-\frac{{M^{2}_{W}}\,{s^{2}_{\theta}}\,
      ( 3 - 8\,{s^{2}_{\theta}} ) }{{c^{2
       }_{\theta}}\,{{\epsilon}^{2}}}}
\nonumber \\
&  \hspace{\eqnoffst} -{\frac{{M^{2}_{W}}\,{s^{2}_{\theta}}\,
      ( 3 - 8\,{s^{2}_{\theta}} ) }{6\,
      {c^{2}_{\theta}}\,\epsilon}}\{11
 +12\,{\ln{c^2_{\theta} }}
-12\,{\ln{M^{2}_{W}}}\}
\nonumber\\
\intertext{\bf Diagram (f)}
\intertext{Internal photon}
\Pi_{WW}^{(2\,{\rm f})}(0)=&-M_W^2\frac{16}{3}s_\theta^4
                           \frac{(n-2)}{n}I_0(1,1,1,1,n,M_W^2)\\
\Delta\Pi_{WW}^{(2\,{\rm f})}(q^2)=& {-\frac{4\,{M^{2}_{W}}\,{s^{4}_{\theta}}}
    {3\,{{\epsilon}^{2}}}}
\nonumber \\
&  \hspace{\eqnoffstc} {-\frac{{M^{2}_{W}}\,{s^{4}_{\theta}}}
    {9\,\epsilon\,{q^4}}}\Big\{8\,{M^{2}_{W}}\,{q^2}
+58\,{q^4}
+ [ 8\,{M^{4}_{W}} +
      32\,{M^{2}_{W}}\,{q^2} ] \,
    {\ln{M^{2}_{W}}}
\nonumber \\
&  \hspace{\eqnoffst} -[ 8\,{M^{4}_{W}} + 32\,{M^{2}_{W}}\,{q^2} +
     24\,{q^4} ] \,{\ln({q^2}+{M^{2}_{W}})}\Big\}
\nonumber\\
\intertext{Internal $Z^0$}
\Pi_{WW}^{(2\,{\rm f})}(0)=&-2M_W^2\frac{s_\theta^4}{c_\theta^4}
                       \left(1-2s_\theta^2+\frac{8}{3}s_\theta^4\right)
                       \frac{(n-2)}{n}I_2(-1,n,M_W^2,M_Z^2)\\
\Delta\Pi_{WW}^{(2\,{\rm f})}(q^2)=&  {-\frac{{M^{2}_{W}}\,{s^{4}_{\theta}}\,
      ( 3 - 6\,{s^{2}_{\theta}} +
        8\,{s^{4}_{\theta}} ) }{6\,
      {c^{4}_{\theta}}\,{{\epsilon}^{2}}}}
\nonumber \\
&  \hspace{\eqnoffstc} +{\frac{{M^{2}_{W}}\,{s^{4}_{\theta}}\,
      ( 3 - 6\,{s^{2}_{\theta}} +
        8\,{s^{4}_{\theta}} ) }{\epsilon\,
      ( 72\,{c^{8}_{\theta}}\,
         ( 1 + {c^{2}_{\theta}} ) \,
         {M^{2}_{W}}\,{q^4} +
        36\,{c^{10}_{\theta}}\,{q^6} +
        36\,{c^{6}_{\theta}}\,{M^{4}_{W}}\,{q^2}\,
         {s^{4}_{\theta}} ) }}
\nonumber \\
&  \hspace{\eqnoffsta}\times \Big\{
  -5\,{c^{6}_{\theta}}\,{q^6}
 -4\,{M^{6}_{W}}\,{s^{4}_{\theta}}\,
  ( {\ln{c^2_{\theta} }} + {s^{2}_{\theta}} )
 +6\,{c^{4}_{\theta}}\,{M^{2}_{W}}\,{q^4}\,
  ( 2 + 2\,{\ln{c^2_{\theta} }} +
    {s^{2}_{\theta}} )
\nonumber \\
&  \hspace{\eqnoffst} +{c^{2}_{\theta}}\,{M^{4}_{W}}\,{q^2}\,
  ( 8\,{\ln{c^2_{\theta} }}\,
     ( 2 - {s^{2}_{\theta}} )  +
    {s^{2}_{\theta}}\,
     ( 16 + 3\,{s^{2}_{\theta}} )  )
\nonumber \\
&  \hspace{\eqnoffst} - [ 8\,{c^{2}_{\theta}}\,{M^{4}_{W}}\,
       {q^2}\,{s^{4}_{\theta}} -
      4\,{M^{6}_{W}}\,{s^{6}_{\theta}} +
      4\,{c^{4}_{\theta}}\,{M^{2}_{W}}\,{q^4}\,
       ( 4 - {s^{2}_{\theta}} ) ] \,
    {\ln{M^{2}_{W}}}
\nonumber \\
&  \hspace{\eqnoffst} -[ 12\,{c^{6}_{\theta}}\,{q^6} +
     20\,{c^{2}_{\theta}}\,{M^{4}_{W}}\,{q^2}\,
      {s^{4}_{\theta}} -
     4\,{M^{6}_{W}}\,{s^{6}_{\theta}}
\nonumber \\
&  \hspace{\eqnoffstb} +
     4\,{c^{4}_{\theta}}\,{M^{2}_{W}}\,{q^4}\,
      ( 16 - 7\,{s^{2}_{\theta}} ) ] \,
   {{\tilde B_0}({q^2},{M^{2}_{W}},{M^{2}_{Z}}
    )}\Big\}
\nonumber\\
\intertext{Internal $Z$-$\gamma$ mixing}
\Pi_{WW}^{(2\,{\rm f})}(0)=&4\frac{s_\theta^4}{c_\theta^2}
                            \left(1-\frac{8}{3}s_\theta^2\right)
                            \frac{(n-2)}{n}I_1(0,n,M_W^2,M_Z^2)\\
\Delta\Pi_{WW}^{(2\,{\rm f})}(q^2)=&  {\frac{{M^{2}_{W}}\,{s^{4}_{\theta}}\,
      ( 3 - 8\,{s^{2}_{\theta}} ) }{3\,
      {c^{2}_{\theta}}\,{{\epsilon}^{2}}}}
\nonumber \\
&  \hspace{\eqnoffstc} +{\frac{{s^{2}_{\theta}}\,
      ( 3 - 8\,{s^{2}_{\theta}} ) }{54\,
      {c^{6}_{\theta}}\,\epsilon\,
      {q^4}}} \Big\{ 8\,{c^{6}_{\theta}}\,{q^6}\,
   {s^{2}_{\theta}}
 + {c^{4}_{\theta}}\,{M^{2}_{W}}\,{q^4}\,
    {s^{2}_{\theta}}\,
    ( 19 - 4\,{\ln{c^2_{\theta} }} -
      16\,{s^{2}_{\theta}} )
\nonumber \\
&  \hspace{\eqnoffst} +4\,{c^{2}_{\theta}}\,{M^{4}_{W}}\,{q^2}\,
  {s^{2}_{\theta}}\,
  ( 1 - ( 2 - {\ln{c^2_{\theta} }} ) \,
     {s^{2}_{\theta}} + 2\,{s^{4}_{\theta}} )
\nonumber \\
&  \hspace{\eqnoffst} + [ 4\,{c^{6}_{\theta}}\,{M^{6}_{W}}\,
       {s^{2}_{\theta}} +
      4\,{c^{4}_{\theta}}\,{M^{2}_{W}}\,{q^4}\,
       {s^{2}_{\theta}}\,
       ( 2 - {s^{2}_{\theta}} )
\nonumber \\
&  \hspace{\eqnoffstb}+   4\,{M^{4}_{W}}\,{q^2}\,{s^{2}_{\theta}}\,
       ( 3 - 9\,{s^{2}_{\theta}} +
         8\,{s^{4}_{\theta}} - 2\,{s^{6}_{\theta}}
          )  ] \,{\ln{M^{2}_{W}}}
\nonumber \\
&  \hspace{\eqnoffst} -[ 4\,{c^{6}_{\theta}}\,{M^{6}_{W}}\,
     {s^{2}_{\theta}} +
    12\,{c^{6}_{\theta}}\,{M^{4}_{W}}\,{q^2}\,
     {s^{2}_{\theta}} +
    12\,{c^{6}_{\theta}}\,{M^{2}_{W}}\,{q^4}\,
     {s^{2}_{\theta}} +
    4\,{c^{6}_{\theta}}\,{q^6}\,{s^{2}_{\theta}}
     ] \,{\ln({q^2}+{M^{2}_{W}})}
\nonumber \\
&  \hspace{\eqnoffst} - [ 4\,{c^{6}_{\theta}}\,{q^6}\,
        {s^{2}_{\theta}} +
       4\,{c^{2}_{\theta}}\,{M^{4}_{W}}\,{q^2}\,
        {s^{6}_{\theta}} -
       8\,{c^{4}_{\theta}}\,{M^{2}_{W}}\,{q^4}\,
        {s^{2}_{\theta}}\,
        ( 4 + {s^{2}_{\theta}} )  ] \,
     {{\tilde B_0}({q^2},{M^{2}_{W}},{M^{2}_{Z}}
      )} \Big\}
\nonumber\\
\intertext{\bf Diagrams (g), (h), (i) and (j) }
\intertext{Internal photon}
\Pi_{WW}^{(2\,{\rm g+h+i+j})}(0)=&8s_\theta^2\frac{(n-2)}{n}
         \{ 2I_0( 0, 1, 1, 1, n, M_W^2 )
          -  I_0( 1, 1, 2, -1, n, M_W^2 )
\nonumber \\ & \qquad\qquad\qquad\qquad
          -  I_0( -1, 1, 2, 1, n, M_W^2 )
         \}\\
\Delta\Pi_{WW}^{(2\,{\rm g+h+i+j})}(q^2)=& {-\frac{{s^{2}_{\theta}}}{9\,{{\epsilon}^{2}}}}\{
  54\,{M^{2}_{W}} - 2\,{q^2}\}
\nonumber \\
&  \hspace{\eqnoffstc} -{\frac{{s^{2}_{\theta}}}{27\,\epsilon\,{q^4}}}\Big\{
  60\,{M^{4}_{W}}\,{q^2}
 +519\,{M^{2}_{W}}\,{q^4}
 +59\,{q^6}
\nonumber \\
&  \hspace{\eqnoffst} + [ 60\,{M^{6}_{W}} +
      252\,{M^{4}_{W}}\,{q^2} -
      360\,{M^{2}_{W}}\,{q^4} ] \,
    {\ln{M^{2}_{W}}}
 -216\,{q^6}\,{\ln{q^2}}
\nonumber \\
&  \hspace{\eqnoffst} -[ 60\,{M^{6}_{W}} + 252\,{M^{4}_{W}}\,{q^2} -
     36\,{M^{2}_{W}}\,{q^4} - 228\,{q^6} ] \,
   {\ln({q^2} + {M^{2}_{W}})}\Big\}
\nonumber\\
\intertext{Internal $Z^0$}
\Pi_{WW}^{(2\,{\rm g+h+i+j})}(0)=&
         \frac{8c_\theta^4}{M_W^2 s_\theta^2}\frac{(n-2)}{n}
         \{ I_0( 0, 1, 2, -1, n, M_Z^2 )
          - I_0( 0, 1, 2, -1, n, M_W^2 )
\nonumber \\ & \qquad\qquad
          + I_0( -2, 1, 2, 1, n, M_Z^2 )
          - I_0( -2, 1, 2, 1, n, M_W^2 )
          \\ & \qquad\qquad
      - 2 I_0( -1, 1, 1, 1, n, M_Z^2 )
      + 2 I_0( -1, 1, 1, 1, n, M_W^2 ) \}
\nonumber\\
\Delta\Pi_{WW}^{(2\,{\rm g+h+i+j})}(q^2)=&{\frac{1}{9\,{{\epsilon}^{2}}}}\{
  {q^2}\,( 2 - 2\,{s^{2}_{\theta}} )  -
   54\,{M^{2}_{W}}\,
    ( 2 - {s^{2}_{\theta}} ) \}
\nonumber \\
&  \hspace{\eqnoffstc} -{\frac{1}{27\,{c^{2}_{\theta}}\,\epsilon\,{q^2}}}
\Big\{  515\,{c^{4}_{\theta}}\,{q^4}
-60\,{M^{4}_{W}}\,{s^{2}_{\theta}}\,
  ( {\ln{c^2_{\theta} }} + {s^{2}_{\theta}}
     )
+3\,{c^{2}_{\theta}}\,{M^{2}_{W}}\,{q^2}\,
  ( 44\,{\ln{c^2_{\theta} }} +
    45\,( 2 - {s^{2}_{\theta}} )  )
\nonumber \\
&  \hspace{\eqnoffst}+ [ 60\,{M^{4}_{W}}\,{s^{4}_{\theta}} -
      132\,{M^{2}_{W}}\,{q^2}\,
       ( 2 - 3\,{s^{2}_{\theta}} +
         {s^{4}_{\theta}} ) ] \,
    {\ln{M^{2}_{W}}}
-216\,{c^{4}_{\theta}}\,{q^4}\,{\ln{q^2}}
\nonumber \\
&  \hspace{\eqnoffst} - [ 228\,{c^{4}_{\theta}}\,{q^4} -
       60\,{M^{4}_{W}}\,{s^{4}_{\theta}} -
       192\,{M^{2}_{W}}\,{q^2}\,
        ( 2 - 3\,{s^{2}_{\theta}} +
          {s^{4}_{\theta}} ) ] \,
     {{\tilde B_0}({q^2},{M^{2}_{W}},{M^{2}_{Z}}
      )} \Big\}
\nonumber \\
\intertext{\bf Diagrams (k) and (l) }
\Pi_{WW}^{(2\,{\rm k+l})}(0)=&2\frac{(n-2)^2}{n}
         \{ I_0( -1, 1, 2, 1, n, M_W^2 )
          - I_0( 0, 1, 1, 1, n,  M_W^2 )
         \}\\
\Delta\Pi_{WW}^{(2\,{\rm k+l})}(q^2)=& {\frac{1}{3\,{{\epsilon}^{2}}}}\{6\,{M^{2}_{W}} + 2\,{q^2}
  \}
+{\frac{1}{\epsilon}}\{{M^{2}_{W}}
+{\frac{8\,{q^2}}{3}}
-4\,{M^{2}_{W}}\,{\ln{M^{2}_{W}}}
-{\frac{4\,{q^2}\,{\ln{q^2}}}{3}}\}
\nonumber\\
\intertext{\bf Diagram (m)}
\intertext{Internal photon}
\Pi_{WW}^{(2\,{\rm m})}(0)=&0\\
\Delta\Pi_{WW}^{(2\,{\rm m})}(q^2)=& {\frac{2\,{q^2}\,{s^{2}_{\theta}}}{9\,{{\epsilon}^{2}}}}
+{\frac{{q^2}\,{s^{2}_{\theta}}}{9\,\epsilon}}\{11
-4\,{\ln{q^2}}\}
\nonumber  \\
\intertext{Internal $Z^0$}
\Pi_{WW}^{(2\,{\rm m})}(0)=&-\frac{1}{c_\theta^2}
         \left(1-2s_\theta^2+\frac{2}{3}s_\theta^4\right)
         \frac{(n-2)}{n}
           \{ (n-2) I_0( 0, 1, 1, 1, n, M_Z^2 )
\nonumber \\ &  \qquad\qquad\qquad\qquad\qquad\qquad\qquad
                 -4 I_0( -1, 1, 2, 1, n, M_Z^2 )
          \\ &  \qquad\qquad\qquad\qquad\qquad\qquad\qquad
                  + I_0( -2, 1, 2, 2, n, M_Z^2 ) \}\nonumber\\
\Delta\Pi_{WW}^{(2\,{\rm m})}(q^2)=& {\frac{(3 - 6\,{s^{2}_{\theta}} + 2\,{s^{4}_{\theta}})}
    {9\,{c^{4}_{\theta}}\,{{\epsilon}^{2}}}}\{
  3\,{M^{2}_{W}} + {c^{2}_{\theta}}\,{q^2}\}
\nonumber \\
&  \hspace{\eqnoffstc} +{\frac{(3 - 6\,{s^{2}_{\theta}} + 2\,{s^{4}_{\theta}})}
    {18\,{c^{4}_{\theta}}\,\epsilon}}\Big\{
  3\,( 1 + 4\,{\ln{c^2_{\theta} }} ) \,
   {M^{2}_{W}}
+11\,{c^{2}_{\theta}}\,{q^2}
-12\,{M^{2}_{W}}\,{\ln{M^{2}_{W}}}
-4\,{c^{2}_{\theta}}\,{q^2}\,{\ln{q^2}}\Big\}
\nonumber \\
\intertext{\bf Diagrams (n) and (o) }
\intertext{Internal photon}
\Pi_{WW}^{(2\,{\rm n+o})}(0)=&0\\
\Delta\Pi_{WW}^{(2\,{\rm n+o})}(q^2)=& {\frac{4\,{q^2}\,{s^{2}_{\theta}}}{9\,{{\epsilon}^{2}}}}
+{\frac{{q^2}\,{s^{2}_{\theta}}}{9\,\epsilon}}\{16
-8\,{\ln{q^2}}\}
\nonumber \\
\intertext{Internal $Z^0$}
\Pi_{WW}^{(2\,{\rm n+o})}(0)=&\frac{1}{c_\theta^2}
         \left(1-2s_\theta^2+\frac{4}{3}s_\theta^4\right)
         \frac{(n-2)^2}{n}
         \{ I_0( -1, 1, 2, 1, n, M_Z^2 )
\nonumber \\ & \qquad\qquad\qquad\qquad\qquad\qquad
          - I_0( 0, 1, 1, 1, n,  M_Z^2 )
         \}\\
\Delta\Pi_{WW}^{(2\,{\rm n+o})}(q^2)=& {\frac{(3 - 6\,{s^{2}_{\theta}} + 4\,{s^{4}_{\theta}})}
    {9\,{c^{4}_{\theta}}\,{{\epsilon}^{2}}}}\{
  3\,{M^{2}_{W}} + {c^{2}_{\theta}}\,{q^2}\}
\nonumber \\
&  \hspace{\eqnoffstc} +{\frac{(3 - 6\,{s^{2}_{\theta}} + 4\,{s^{4}_{\theta}})}
    {18\,{c^{4}_{\theta}}\,\epsilon}
    }\Big\{3\,
   ( 1 + 4\,{\ln{c^2_{\theta} }} ) \,
   {M^{2}_{W}}
+8\,{c^{2}_{\theta}}\,{q^2}
-12\,{M^{2}_{W}}\,
  {\ln{M^{2}_{W}}}
-4\,{c^{2}_{\theta}}\,{q^2}\,{\ln{q^2}}\Big\}
\nonumber\\
\end{align}

\begin{flushleft}
{\bf Tadpole Diagrams}
\end{flushleft}

For completeness the contributions of the ${\cal O}(N_f\alpha^2)$
tadpole diagrams, Fig.\ref{fig:WTadpole}, are given. As above
an overall factor of $\left(g^2/(16\pi)^2\right)^2\delta_{\mu\nu}$
has been omitted from all diagrams.

\begin{align}
\intertext{{\bf Diagram (a)}}
\Pi_{WW}^{(2\,{\rm a})}(0)=&-2c_h^2
                            (n-2)I_0(-1,2,1,1,n,M_W^2)\\
\intertext{\bf Diagram (b)}
\Pi_{WW}^{(2\,{\rm b})}(0)=&-2\frac{M_W^2}{c_h^2 c_\theta^2}
              \left(1-2s_\theta^2+\frac{8}{3}s_\theta^4\right)
              (n-2)I_0(-1,2,1,1,n,M_W^2)
\end{align}

\section{Vertex Corrections}

In this appendix the vertex and external leg corrections for massless
on-shell fermions coupling to the $W$ boson are given.
These take general form
\[
i\frac{g}{\sqrt{2}}\gamma_\mu\gamma_L V^{(2)}(0)
\]
with the contribution of the various diagrams to $V^{(2)}(0)$ being
given below.
Their net effect on the inverse muon lifetime is to induce a shift of
$\Delta\Gamma^{(2)}=4 \Gamma^{(0)} V^{(2)}(0)$
or equivalently produce a contribution of $2V^{(2)}(0)$ to
$\Delta r^{(2)}$.

The diagrams are labeled according to Fig.\ref{fig:VertexDiags}.
An overall factor of $\left(g^2/(16\pi)^2\right)^2$ has been
omitted from all diagrams.

\begin{align}
\intertext{\bf Diagrams (a), (b) and (c) }
\intertext{Internal photon}
V^{(2\,{\rm a+b+c})}(0)=&\frac{8}{3}s_\theta^4
                    \frac{(n-2)(n-4)}{n}I_0(1,1,1,1,n,\Lambda^2)\\
\intertext{Internal $Z^0$}
V^{(2\,{\rm a+b+c})}(0)=&\left(1-2s_\theta^2+\frac{8}{3}s_\theta^4\right)
                    \frac{(n-2)(n-4)}{n}I_0(0,2,1,1,n,M_Z^2)\\
\intertext{Internal $Z$-$\gamma$ mixing}
V^{(2\,{\rm a+b+c})}(0)=&2s_\theta^2\left(1-\frac{8}{3}s_\theta^2\right)
                    \frac{(n-2)(n-4)}{n}I_0(1,1,1,1,n,M_Z^2)\\
\intertext{\bf Diagrams (d) and (e) }
V^{(2\,{\rm d+e})}(0)=&\frac{(n-2)(n-4)}{n}I_0(0,2,1,1,n,M_W^2)\\
\intertext{\bf Diagrams (f) and (g) }
\intertext{Internal photon}
V^{(2\,{\rm f+g})}(0)=&-16 s_\theta^4\frac{(n-2)}{n}I_0(1,1,1,1,n,M_W^2)\\
\intertext{Internal $Z^0$}
V^{(2\,{\rm f+g})}(0)=&-6\left(1-2s_\theta^2+\frac{8}{3}s_\theta^4\right)
                     \frac{(n-2)}{n}I_2(-1,n,M_W^2,M_Z^2)\\
\intertext{Internal $Z$-$\gamma$ mixing}
V^{(2\,{\rm f+g})}(0)=&-12\left(1-\frac{8}{3}s_\theta^2\right)
                        \frac{(n-2)}{n}I_1(0,n,M_W^2,M_Z^2)\\
\intertext{\bf Diagrams (h) and (i) }
\intertext{Internal photon}
V^{(2\,{\rm h+i})}(0)=&-6s_\theta^2\frac{(n-2)}{n}I_0(0,2,1,1,n,M_W^2)\\
\intertext{Internal $Z^0$}
V^{(2\,{\rm h+i})}(0)=&-6c_\theta^2\frac{(n-2)}{n}I_2(-1,n,M_Z^2,M_W^2)\\
\intertext{\bf Diagrams (j), (k), (l) and (m) }
\intertext{Internal photon}
V^{(2\,{\rm j+k+l+m})}(0)=&2 s_\theta^2 \frac{(n-2)}{n}
        \{ 3 I_0(1,1,1,1,n,M_W^2) - I_0(2,1,2,-1,n,M_W^2)
        \nonumber\\ &  \qquad\qquad\qquad\qquad\qquad
          - I_0(0,1,2,1,n,M_W^2) \}
\intertext{Internal $Z^0$}
V^{(2\,{\rm j+k+l+m})}(0)=&\frac{2c_\theta^4}{M_W^2 s_\theta^2}
        \frac{(n-2)}{n}
        \{ 3I_0(0,1,1,1,n,M_W^2) - 3I_0(0,1,1,1,n,M_Z^2 )
        \nonumber\\ & \qquad\qquad
           -I_0(1,1,2,-1,n,M_W^2)+I_0(1,1,2,-1,n,M_Z^2)
                 \\ & \qquad\qquad
           -I_0(-1,1,2,1,n,M_W^2)+I_0(-1,1,2,1,n,M_Z^2)
         \}\nonumber
\end{align}

\section{Box Diagrams}

In this appendix expressions are given for the box diagrams of
Fig.\ref{fig:BoxDiagrams}. The methods used in their calculation
have been discussed in section~\ref{sect:BoxDiagrams}. All
diagrams are simply proportional to the tree-level muon decay
amplitude. Their net effect on the inverse muon lifetime
is to induce a shift of $\Delta\Gamma^{(2)}=2 \Gamma^{(0)} B^{(2)}$
or equivalently produce a contribution of $B$ to $\Delta r^{(2)}$.
The diagrams are labeled according to Fig.\ref{fig:BoxDiagrams}
An overall factor of $\left(g^2/(16\pi)^2\right)^2$ has been
omitted from all diagrams.

\begin{align}
\intertext{\bf Diagrams (a) and (b) }
\intertext{Internal $Z^0$}
B^{(2\,{\rm a+b})}=&-\frac{2}{c_\theta^4}(1-2s_\theta^2)
                     \left(1-2s_\theta^2+\frac{8}{3}s_\theta^4\right)
                     \frac{(n-2)}{n}I_1(0,n,M_W^2,M_Z^2)\\
\intertext{Internal $Z$-$\gamma$ mixing}
B^{(2\,{\rm a+b})}(0)=&-4\frac{s_\theta^2}{c_\theta^2}
                     \left(1-\frac{8}{3}s_\theta^2\right)
                     \frac{(n-2)}{n}I_1(1,n,M_W^2,M_Z^2)\\
\intertext{\bf Diagrams (c) and (d) }
B^{(2\,{\rm c+d})}=&-\frac{2}{c_\theta^2}(1-2s_\theta^2)
                     \frac{(n-2)}{n}I_2(0,n,M_Z^2,M_W^2)
\end{align}

\section{Counterterm Insertions}
\label{sect:CTInsert}

For an internal line in a Feynman diagram, representing a physical
particle, there is a cancellation of wavefunction counterterms between
those in the 2-point counterterm and those in vertices at its endpoints.
While this happens trivially for the $W$ boson, the cancellation is
less straightforward in the case of neutral bosons where $Z$-$\gamma$
mixing is a complicating factor.

In this appendix, identities, valid to ${\cal O}(\alpha)$,
are given for the counterterm insertions on internal neutral boson
lines. The cancellation of the wavefunction
counterterms, $\delta Z_W^{(1)}$ and $\delta Z_B^{(1)}$, has been
explicitly carried out. Note that wavefunction counterterms associated
with the external particles have not been included and must be taken
into account separately. A similar cancellation of wavefunction
counterterms in the presence of $Z$-$\gamma$ mixing was carried out
in ref.\cite{ZMass4}.

As in the text the notation $Z$,$\gamma$ means that all
possibilities are to be included. In the identities, $g\beta_L$ and
$g\beta_R$ are the left- and right-handed couplings of the $Z^0$ to
the fermion
\begin{equation}
\beta_{L}=\frac{t_{3}-s_\theta^2 Q}{c_\theta},\ \ \ \ \ \
\beta_{R}=-\frac{s_\theta^2 Q}{c_\theta}
\end{equation}
in which $t_3$ is its weak isospin and $Q$ is its electric charge.
Unprimed quantities are associated with the fermion current
labeled 1 and primed with the current labeled 2. Square brackets
[\ ] indicate that the enclosed quantities pertain to the fermion
current given by the its subscript
\begin{align}
&
\begin{picture}(92,72)(0,0)
\ArrowLine(0,0)(12,36)
\ArrowLine(12,36)(0,72)
\Text(4,36)[r]{1}
\Photon(12,36)(80,36){4}{6}
\Text(46,42)[b]{$Z$,$\gamma$}
\ArrowLine(92,0)(80,36)         \Vertex(80,36){1}
\ArrowLine(80,36)(92,72)
\Text(88,36)[l]{2}
\LongArrow(39,28)(53,28)
\Text(46,25)[t]{$q$}
\SetWidth{1.0}
\Line(5,43)(19,29)
\Line(5,29)(19,43)
\end{picture}
\qquad\raisebox{36pt}{+}\qquad
\begin{picture}(92,72)(0,0)
\ArrowLine(0,0)(12,36)          \Vertex(12,36){1}
\ArrowLine(12,36)(0,72)
\Text(4,36)[r]{1}
\Photon(12,36)(80,36){4}{6}
\Text(26,42)[b]{$Z$,$\gamma$}   \Text(66,42)[b]{$Z$,$\gamma$}
\ArrowLine(92,0)(80,36)         \Vertex(80,36){1}
\ArrowLine(80,36)(92,72)
\Text(88,36)[l]{2}
\SetWidth{1.0}
\Line(39,43)(53,29)
\Line(39,29)(53,43)
\end{picture}
\qquad\raisebox{36pt}{+}\qquad
\begin{picture}(92,72)(0,0)
\ArrowLine(0,0)(12,36)          \Vertex(12,36){1}
\ArrowLine(12,36)(0,72)
\Text(4,36)[r]{1}
\Photon(12,36)(80,36){4}{6}
\Text(46,42)[b]{$Z$,$\gamma$}
\ArrowLine(92,0)(80,36)
\ArrowLine(80,36)(92,72)
\Text(88,36)[l]{2}
\SetWidth{1.0}
\Line(73,43)(87,29)
\Line(73,29)(87,43)
\end{picture}\notag\\
&\ \ \ \ =\frac{2}{q^2}\left(s_\theta^2\frac{\delta g^{(1)}}{g}
                   +c_\theta^2\frac{\delta g^{\prime(1)}}{g^\prime}
              \right)
 [igs_\theta Q\gamma_\mu]_1 [igs_\theta Q^\prime\gamma_\mu]_2\notag\\
&\ \ \ \ +\frac{2s_\theta c_\theta}{q^2+M_Z^2}
              \left(\frac{\delta g^{(1)}}{g}
                   -\frac{\delta g^{\prime(1)}}{g^\prime}
              \right)
 [igs_\theta Q\gamma_\mu]_1
 [ig\gamma_\mu(\beta_L^\prime\gamma_L+\beta_R^\prime\gamma_R)]_2\notag\\
&\ \ \ \ +\frac{2s_\theta c_\theta}{q^2+M_Z^2}
              \left(\frac{\delta g^{(1)}}{g}
                   -\frac{\delta g^{\prime(1)}}{g^\prime}
              \right)
 [ig\gamma_\mu(\beta_L\gamma_L+\beta_R\gamma_R)]_1
 [igs_\theta Q^\prime\gamma_\mu]_2\\
&\ \ \ \ +\frac{2}{q^2+M_Z^2}\left(c_\theta^2\frac{\delta g^{(1)}}{g}
                         +s_\theta^2\frac{\delta g^{\prime(1)}}{g^\prime}
              \right)
 [ig\gamma_\mu(\beta_L\gamma_L+\beta_R\gamma_R)]_1
 [ig\gamma_\mu(\beta_L^\prime\gamma_L+\beta_R^\prime\gamma_R)]_2\notag\\
&\ \ \ \ -\frac{\delta M_Z^{2(1)}}{(q^2+M_Z^2)^2}
 [ig\gamma_\mu(\beta_L\gamma_L+\beta_R\gamma_R)]_1
 [ig\gamma_\mu(\beta_L^\prime\gamma_L+\beta_R^\prime\gamma_R)]_2\notag
\end{align}

\begin{align}
&
\begin{picture}(92,72)(0,0)
\ArrowLine(0,0)(12,36)
\ArrowLine(12,36)(0,72)
\Text(4,36)[r]{1}
\Photon(12,36)(80,36){4}{6}
\Text(46,42)[b]{$Z$,$\gamma$}
\Photon(92,0)(80,36){-4}{3.5}          \Vertex(80,36){1}
\Text(85,0)[br]{$W$}
\Photon(80,36)(92,72){4}{3.5}
\Text(85,72)[tr]{$W$}
\LongArrow(39,28)(53,28)
\Text(46,25)[t]{$q$}
\SetWidth{1.0}
\Line(5,43)(19,29)
\Line(5,29)(19,43)
\end{picture}
\qquad\raisebox{36pt}{+}\qquad
\begin{picture}(92,72)(0,0)
\ArrowLine(0,0)(12,36)          \Vertex(12,36){1}
\ArrowLine(12,36)(0,72)
\Text(4,36)[r]{1}
\Photon(12,36)(80,36){4}{6}
\Text(26,42)[b]{$Z$,$\gamma$}   \Text(66,42)[b]{$Z$,$\gamma$}
\Photon(92,0)(80,36){-4}{3.5}   \Vertex(80,36){1}
\Text(85,0)[br]{$W$}
\Text(85,72)[tr]{$W$}
\Photon(80,36)(92,72){4}{3.5}
\SetWidth{1.0}
\Line(39,43)(53,29)
\Line(39,29)(53,43)
\end{picture}
\qquad\raisebox{36pt}{+}\qquad
\begin{picture}(92,72)(0,0)
\ArrowLine(0,0)(12,36)          \Vertex(12,36){1}
\ArrowLine(12,36)(0,72)
\Text(4,36)[r]{1}
\Photon(12,36)(80,36){4}{6}
\Text(46,42)[b]{$Z$,$\gamma$}
\Photon(92,0)(80,36){-4}{3.5}
\Text(85,0)[br]{$W$}
\Photon(80,36)(92,72){4}{3.5}
\Text(85,72)[tr]{$W$}
\SetWidth{1.0}
\Line(73,43)(87,29)
\Line(73,29)(87,43)
\end{picture}\notag\\
&\qquad\qquad\qquad\qquad
=\frac{2}{q^2}\left(s_\theta^2\frac{\delta g^{(1)}}{g}
                   +c_\theta^2\frac{\delta g^{\prime(1)}}{g^\prime}
              \right)
 [igs_\theta Q\gamma_\mu]_1\ (gs_\theta)\notag\\
&\qquad\qquad\qquad\qquad
+\frac{2s_\theta c_\theta}{q^2+M_Z^2}
              \left(\frac{\delta g^{(1)}}{g}
                   -\frac{\delta g^{\prime(1)}}{g^\prime}
              \right)
 [igs_\theta Q\gamma_\mu]_1\ (gc_\theta)\\
&\qquad\qquad\qquad\qquad
+\frac{2}{q^2+M_Z^2}
              \left(\frac{\delta g^{(1)}}{g}\right)
 [ig\gamma_\mu(\beta_L\gamma_L+\beta_R\gamma_R)]_1\ (gc_\theta)\notag\\
&\qquad\qquad\qquad\qquad
-\frac{\delta M_Z^{2(1)}}{(q^2+M_Z^2)^2}
 [ig\gamma_\mu(\beta_L\gamma_L+\beta_R\gamma_R)]_1
 \ (gc_\theta)\notag
\end{align}

\begin{align}
&
\begin{picture}(92,72)(0,0)
\Photon(0,0)(12,36){-4}{3.5}
\Text(7,0)[bl]{$W$}
\Photon(12,36)(0,72){4}{3.5}
\Text(7,72)[tl]{$W$}
\Photon(12,36)(80,36){4}{6}
\Text(46,42)[b]{$Z$,$\gamma$}
\Photon(92,0)(80,36){-4}{3.5}          \Vertex(80,36){1}
\Text(85,0)[br]{$W$}
\Photon(80,36)(92,72){4}{3.5}
\Text(85,72)[tr]{$W$}
\LongArrow(39,28)(53,28)
\Text(46,25)[t]{$q$}
\SetWidth{1.0}
\Line(5,43)(19,29)
\Line(5,29)(19,43)
\end{picture}
\qquad\raisebox{36pt}{+}\qquad
\begin{picture}(92,72)(0,0)
\Photon(0,0)(12,36){-4}{3.5}          \Vertex(12,36){1}
\Text(7,0)[bl]{$W$}
\Photon(12,36)(0,72){4}{3.5}
\Text(7,72)[tl]{$W$}
\Photon(12,36)(80,36){4}{6}
\Text(26,42)[b]{$Z$,$\gamma$}   \Text(66,42)[b]{$Z$,$\gamma$}
\Photon(92,0)(80,36){-4}{3.5}   \Vertex(80,36){1}
\Text(85,0)[br]{$W$}
\Text(85,72)[tr]{$W$}
\Photon(80,36)(92,72){4}{3.5}
\SetWidth{1.0}
\Line(39,43)(53,29)
\Line(39,29)(53,43)
\end{picture}
\qquad\raisebox{36pt}{+}\qquad
\begin{picture}(92,72)(0,0)
\Photon(0,0)(12,36){-4}{3.5}          \Vertex(12,36){1}
\Text(7,0)[bl]{$W$}
\Photon(12,36)(0,72){4}{3.5}
\Text(7,72)[tl]{$W$}
\Photon(12,36)(80,36){4}{6}
\Text(46,42)[b]{$Z$,$\gamma$}
\Photon(92,0)(80,36){-4}{3.5}
\Text(85,0)[br]{$W$}
\Photon(80,36)(92,72){4}{3.5}
\Text(85,72)[tr]{$W$}
\SetWidth{1.0}
\Line(73,43)(87,29)
\Line(73,29)(87,43)
\end{picture}\notag\\
&\qquad\qquad\qquad\qquad
=\frac{2}{q^2}\left(s_\theta^2\frac{\delta g^{(1)}}{g}
                   +c_\theta^2\frac{\delta g^{\prime(1)}}{g^\prime}
              \right)
 (gs_\theta)\ (gs_\theta)\notag\\
&\qquad\qquad\qquad\qquad
+\frac{2s_\theta c_\theta}{q^2+M_Z^2}
              \left(\frac{\delta g^{(1)}}{g}
                   -\frac{\delta g^{\prime(1)}}{g^\prime}
              \right)
 (gs_\theta)\ (gc_\theta)\\
&\qquad\qquad\qquad\qquad
+\frac{2}{q^2+M_Z^2}
              \left(\frac{\delta g^{(1)}}{g}\right)
 (gc_\theta)\ (gc_\theta)\notag\\
&\qquad\qquad\qquad\qquad
-\frac{\delta M_Z^{2(1)}}{(q^2+M_Z^2)^2}
 (gc_\theta)\ (gc_\theta)\notag
\end{align}

\end{document}